# Review: Topology of solitons and singular defects in chiral liquid crystals

Jin-Sheng Wu[1] and Ivan I. Smalyukh[1,2,3*]

*[1]Department of Physics and Chemical Physics Program, University of Colorado, Boulder, CO 80309, USA*

*[2]Department of Electrical, Computer and Energy Engineering and Materials Science and Engineering Program, University of Colorado, Boulder, CO 80309, USA*

*[3]Renewable and Sustainable Energy Institute, National Renewable Energy Laboratory and University of Colorado, Boulder, CO 80309, USA*

*\*Email: ivan.smalyukh@colorado.edu*

# Review: Topology of solitons and singular defects in chiral liquid crystals

**Abstract:** Widely known for their uses in displays and electro-optics, liquid crystals are more than just technological marvels. They vividly reveal the topology and structure of various solitonic and singular field configurations, often markedly resembling the ones arising in many field theories and in the areas ranging from particle physics to optics, hard condensed matter and cosmology. In this review, we focus on chiral nematic liquid crystals to show how these experimentally highly accessible systems provide valuable insights into the structure and behavior of fractional, full, and multi-integer two-dimensional skyrmions, dislocations and both abelian and non-abelian defect lines, as well as various three-dimensionally localized, often knotted structures that include hopfions, heliknotons, torons and twistions. We provide comparisons of some of these field configurations with their topological counterparts in chiral magnets, discussing close analogies between these two condensed matter systems.

## 1. Introduction

Being the first anisotropic fluids to be experimentally discovered over 150 years ago [1,2], chiral nematic liquid crystals (LCs) offer a particularly diverse variety of topologically nontrivial field configurations [3]. Uniquely, they serve as a testbed for theories describing singular defects and solitons in chiral magnets and various lamellar systems, as well as topological counterparts of solitonic structures found in particle physics and cosmology and non-abelian defects in low-symmetry ordered media [3–6]. Starting from 60-ies of the 20es century, the foundations for fundamental understanding of topological structures in chiral LCs were introduced by key early works of Kleman and colleagues [3,7–26]. Decades of active research that followed revealed a rich variety of emergent and fascinating phenomena, with implications for the understanding of the inner workings of our World, well beyond LCs and condensed matter and often well ahead of similar experimental and theoretical developments in other branches of physics. Being highly accessible experimentally, chiral nematic LCs reveal how chirality can enable diversity and stability of fascinating topological objects, with both fundamental and applied potential.

In this article, we review the rich, multi-faceted nature of topological structures in chiral nematic LCs, as well as compare them to their topological counterparts in chiral magnets and other physical systems. We start from introducing order parameters and different theoretical models for describing energetic costs of perturbing orientational order within these soft matter media, as well as showing how these models, under certain assumptions and within certain limits, exhibit similarities with the ones describing lamellar systems, magnets, and other ordered media (Section 2). We then overview the basics of homotopy theory and its application to chiral nematic LCs under different levels of coarse graining of description (Section 3). Following this, in Section 4, we discuss the structure and experimental observation of various disclinations and dislocations in chiral nematic LCs. How certain types of descriptions of the chiral nematic medium reveal non-abelian properties of the cholesteric line defects is discussed in Section 5. Two-dimensional solitonic structures, referred to as both twist-escaped disclinations and as fractional, full or multi-integer skyrmions, are described in Section 6, whereas Section 7 is devoted to torons that comprise fragments of such skyrmions along with singular point defects, co-embedded in a uniform background. Fully three-dimensional topological solitons, such as hopfions and heliknotons, are described in Section 8 whereas the emergent dynamic behavior of various solitons and defects is revealed in Section 9. Finally, Section 10 provides concluding remarks and perspectives. This review intends to provide a sense of broad relevance and importance of the research efforts on LC defects and topological solitons. The article is not intended to provide an in-depth comprehensive overview of the entirety of relevant literature, but rather to place our own recent work and other related recent studies into a broader context of this field, showing how these studies build on key breakthroughs in the LC research on defects, starting from key contributions of Kleman and colleagues [3,5–27].

## 2. Directors and free energy of chiral liquid crystals

In the description of nonchiral nematic LCs within a mean-field theory, the nonpolar molecular director field **n(r)**, representing the local average orientation of anisotropic (e.g. rodlike) molecules, is commonly used. The energetic costs of elastic deformations away from the ground state in chiral nematic LCs, also referred to as cholesteric LCs (CLCs), can also be fully described in terms of gradients of **n(r)**. However, the twisting nature of the chiral director configurations implies that the spatial geometry of elastic deformations can be more completely characterized by three non-polar, orthonormal director fields [7,8,18,26,28,29] (Fig. 1a): **λ(r)** for molecular director field (equivalent to **n(r)** and interchangeably used along with **n(r)** as a notation within this review), **χ(r)** for the orientation of helical axis, and **τ(r)** being the cross product of the former two, with the exceptions when the tensorial description is necessary [30,31]. A perfect alignment of helical structure of the ground-state CLC (Fig. 1b,c), for example, has the background of **χ** directors uniformly oriented and everywhere perpendicular to the **λ** field twisting around it. The helical pitch $p$, measured along **χ(r)**, is then defined as the distance containing $2\pi$ rotation of **λ(r)** (Fig. 1a), whereas $p/2$ represents the helical CLC quasi-layer thickness corresponding to $\pi$-rotation of the nonpolar **n**≡**λ** director. The **τ** field of such structure, being orthogonal to the other two fields, is like the $\lambda$ field rotated 90° around the helical axis and exhibits the same structure of helical quasi-layers but shifted vertically along $\chi$ by a quarter of pitch as compared to the $\lambda$ field (Fig. 1c). The detailed description of deriving the $\chi$ and $\tau$ fields from an arbitrary $\lambda$-field alignment can be found in Refs. [31,32]. With the introduction of the two directors $\chi(\mathbf{r})$ and $\tau(\mathbf{r})$, in addition to the molecular field $\lambda(\mathbf{r})$, the elastic distortions in CLCs are respectively identified with these directors. Given three director order parameters (instead of one in nonchiral nematic systems), CLC systems can reveal a richer variety of topological structures as compared to nematic LCs. For example, a defect structure can be singular in $\lambda(\mathbf{r})$ and $\chi(\mathbf{r})$ fields but non-singular in $\tau(\mathbf{r})$ field, or show discontinuity in $\chi(\mathbf{r})$ and $\tau(\mathbf{r})$ while being smooth in the molecular director field. Similarly, spatial distortions of different director fields give rise to different but interrelated interpretations of elastic deformations in CLCs, as discussed below.

In the modelling of a CLC, the equilibrium configuration of directors can be found by minimizing the free energy of its continuum representation, which typically takes the form:

$$F = \int f_{el}\, dV + \int f_s\, dS, \qquad (1)$$

where the elastic bulk energy density is integrated over three-dimensional volume occupied by the CLC and the surface anchoring contribution is integrated over the confining two-dimensional surfaces that enclose this volume. For structures with elastic distortions on the scale much smaller than the cholesteric pitch $p$, the effectively nematic-like, weakly twisted CLC systems can be most effectively described by the Frank-Oseen free energy density for the molecular director field [9,23,26,33–35]:

$$f_{el}^{FO} = \frac{K_{11}}{2}(\nabla \cdot \boldsymbol{\lambda})^2 + \frac{K_{22}}{2}(\boldsymbol{\lambda} \cdot (\nabla \times \boldsymbol{\lambda}) + q)^2 + \frac{K_{33}}{2}(\boldsymbol{\lambda} \times (\nabla \times \boldsymbol{\lambda}))^2$$

$$-\frac{K_{24}}{2}\nabla \cdot (\boldsymbol{\lambda}(\nabla \cdot \boldsymbol{\lambda}) + \boldsymbol{\lambda} \times (\nabla \times \boldsymbol{\lambda})), \qquad (2)$$

where $K_{11}$, $K_{22}$, $K_{33}$, and $K_{24}$ are the Frank elastic moduli for splay, twist, bend, and saddle-splay deformations, respectively. The chirality wavevector $q$ is determined by $q=2\pi/p$. The saddle-splay term is often interpreted as "surface-like" elasticity since the divergence theory assures that one can rewrite the volume integral of the divergence term and find only surface contribution.

When the length scale of elastic deformations is much larger than the cholesteric pitch, using distortions in helical axis is typically convenient and sufficient way to describe the CLC system from a coarse-grained perspective. In this case, the elastic properties of the system are similar to those of smectic A and other lamellar LCs, and the free energy is given by the Lubensky-de Gennes coarse-grained model [9,23,33,35,36]:

$$f_{el}^{CG} = \frac{K_1}{2}(\nabla \cdot \boldsymbol{\chi})^2 + \frac{K_3}{2}(\boldsymbol{\chi} \times (\nabla \times \boldsymbol{\chi}))^2 + \bar{K}\nabla \cdot (\boldsymbol{\chi}(\nabla \cdot \boldsymbol{\chi}) + \boldsymbol{\chi} \times (\nabla \times \boldsymbol{\chi})) + \frac{1}{2}B\beta^2, \qquad (3)$$

where $\beta=(h-h_0)/h_0$ is the scaled difference between the actual CLC quasilayer spacing $h$ and its equilibrium value $h_0$ (corresponding to half-pitch of the CLC). In this expression of elastic energy density (Eq. 3), splay, bend, and saddle-splay elastic deformations of $\boldsymbol{\chi}(\mathbf{r})$ field are quantified by means of $K_1$, $K_3$, and $\bar{K}$ elastic constants, respectively, and the energetic costs of dilation/compression of helical quasi-layers are characterized with the Young modulus $B$. Naturally, these deformation modes in the helical axis field are related to the ones described by Frank elastic moduli of the molecular $\boldsymbol{\lambda}(\mathbf{r})$ director field: the splay deformation of $\boldsymbol{\chi}(\mathbf{r})$ is associated with the bending of $\boldsymbol{\lambda}(\mathbf{r})$, and the $B$-term is equivalent to the twist term in the Frank-Oseen model, as stated by the Lubensky–de Gennes relationships: $K_1=3K_{33}/8$; $B=K_{22}(2\pi/p)^2$. The Kats-Lebedev theory additionally gives $K_3=K_{11}K_{33}/2(K_{11}+K_{33})$, derived by neglecting surface-like terms, also connecting the energetic costs of elastic distortions in different directors [33]. Considering relative contributions of different terms in the above expression for the case of large-scale distortions, this coarse-grained description is often further simplified to the form (analogous to the description of elastic deformations in smectic LCs):

$$f_{el}^{CG} = \frac{K_1}{2}(\nabla \cdot \boldsymbol{\chi})^2 + \overline{K}\nabla \cdot (\boldsymbol{\chi}(\nabla \cdot \boldsymbol{\chi}) + \boldsymbol{\chi} \times (\nabla \times \boldsymbol{\chi})) + \frac{1}{2}B\beta^2. \quad (4)$$

In principle, with expressions of elastic free energy in $\boldsymbol{\lambda}(\mathbf{r})$ (Eq. 2) and $\boldsymbol{\chi}(\mathbf{r})$ (Eq. 3), one could also derive Frank-Oseen-like functional for the free energy of distortion modes expressed in terms of $\boldsymbol{\tau}(\mathbf{r})$ and use the relation $\boldsymbol{\tau}=\boldsymbol{\lambda}\times\boldsymbol{\chi}$ to find their connection to the other descriptions of CLC elasticities using vector analysis, though this is not as helpful for practical uses of such models and we will not do this here.

Similar to the elasticity of smectic layers, one can re-express Eq. 4 in terms of the radii of curvature $R_1$, $R_2$ of helical layers and obtain a simpler equation [33,37,38]:

$$f_{el}^{CG} = \frac{K_1}{2}\left(\frac{1}{R_1}+\frac{1}{R_2}\right)^2 + \frac{\overline{K}}{2}\frac{1}{R_1 R_2} + \frac{1}{2}B\beta^2. \quad (5)$$

Within this description, the saddle-splay $\overline{K}$ term has no contribution for translationally invariant structures (one of the curvatures vanishes), such as the ones due to edge dislocations. Considering this, the coarse-grained elastic energy Eq. 4 can also be rewritten by introducing a layer displacement field $u$ for small two-dimensional distortions in the helical quasilayers, which allows one to re-express both the helical director field and free energy density:

$$\boldsymbol{\chi}(\mathbf{r}) = \pm\left\{-\frac{\partial u}{\partial x}\left(1+\frac{\partial u}{\partial z}\right), 0, 1-\frac{1}{2}\left(\frac{\partial u}{\partial x}\right)^2\right\}$$
$$f_{el}^{CG} = \frac{K_1}{2}\left(\frac{\partial^2 u}{\partial x^2}\right)^2 + \frac{1}{2}B\left(\frac{\partial u}{\partial z}-\frac{1}{2}\left(\frac{\partial u}{\partial x}\right)^2\right)^2 \quad (6)$$

where we can see the curvature of helical layers (the second derivative) gives rise to the splay $K_1$ term while the compressibility $B$ term is dominated by the linear displacement of helical quasilayers in the vertical direction.

In the Frank-Oseen energy density (Eq. 2), the saddle-splay contribution as well as the elastic constant anisotropy are sometimes ignored. By re-writing the twist energy term, dropping the $K_{24}$ term and making one-constant approximation $K_{11}=K_{22}=K_{33}$, Eq. 2 can be simplified to a more compact form:

$$f_{el}^{FO} \cong \frac{K_{11}}{2}(\nabla \cdot \boldsymbol{\lambda})^2 + \frac{K_{22}}{2}(\boldsymbol{\lambda} \cdot (\nabla \times \boldsymbol{\lambda}))^2 + \frac{K_{33}}{2}(\boldsymbol{\lambda} \times (\nabla \times \boldsymbol{\lambda}))^2 + qK_{22}\boldsymbol{\lambda} \cdot (\nabla \times \boldsymbol{\lambda})$$

$$= \frac{K}{2}(\nabla \boldsymbol{\lambda})^2 + qK\boldsymbol{\lambda} \cdot (\nabla \times \boldsymbol{\lambda}). \qquad (7)$$

The CLC's free energy functional in Eq. 7 then reduces to a form resembling that of the micromagnetic Hamiltonian for non-centrosymmetric chiral magnets for $A_m=K/2$, $D_m=Kq$ and $\boldsymbol{\lambda}(\mathbf{r}) \rightarrow \mathbf{m}(\mathbf{r})$ [39–41]:

$$f = A_m(\nabla \mathbf{m})^2 + D_m \mathbf{m} \cdot (\nabla \times \mathbf{m}), \qquad (8)$$

where coefficients $A_m$ and $D_m$ describe the effective exchange energy and the Dzyaloshinskii-Moriya coupling constants for magnetic solids. The similarity between energetics of chiral LCs and chiral magnetic materials (Eqs. 7 and 8) suggests that similarities of structures and phenomena can be anticipated for these different physical systems. It also implies that one could use LCs, a purely classical material, to reproduce or simulate some of the phenomena in magnetic material originating from quantum effects. The recent advent of CLC compositions formed by mixtures of rodlike and bent-core molecules allows for "engineering" elastic constant anisotropies [42–44], so that different elastic constants can be designed to be equal or rather different from each other, which further boosts the predictive power of CLCs as a model system.

The surface anchoring free energy, in its simplest form, can be described as the work required to deviate the director from its equilibrium orientation, or easy axis. Adopting a harmonic-like potential, the Rapini-Papoular surface energy density, expressed as polar and azimuthal terms with the help of polar $\theta$ and azimuthal $\phi$ angles describing director orientation, reads [33,45]:

$$f_s = \frac{1}{2} W_p \sin^2(\theta - \theta_e) + \frac{1}{2} W_a \sin^2 \theta_e \sin^2(\phi - \phi_e). \tag{9}$$

However, the angular coordinates of the easy axis $\theta_e$ and $\phi_e$ of the molecular director $\lambda$ orientation are typically not well-suited for the CLCs with strong helical twists. Mimicking the descriptions of bulk free energy of CLCs, the coarse-grained descriptions of the interactions of quasilayered CLC structures with surfaces have been developed too and are described elsewhere [33].

Even more complete description of energetic costs due to perturbations of order within CLCs can be done using the tensorial approach, allowing one to describe both bulk and surface energy costs of CLCs in terms of the Q-tensor order parameter, a 3-by-3 matrix defined by scalar order parameter $S$ and the molecular director field for a uniaxial LC:

$$\mathbf{Q} = \frac{S}{2}(3\lambda \otimes \lambda - \mathbf{I}), \tag{10}$$

with $\otimes$ being the outer product operator and $\mathbf{I}$ is the identity matrix. By quantifying molecular alignment order through $S$, the tensorial expression of bulk free energy density includes a thermotropic contribution that describe the nematic-isotropic transition of LCs:

$$F = \int (f_{el} + f_{thermo})dV + \int f_s dS, \tag{11}$$

where the thermotropic part is often written in the form of Landau-de Gennes expansion of free energy [46–48]:

$$f_{\text{thermo}} = \frac{A_t}{2}\text{Tr}(\mathbf{Q}^2) + \frac{B_t}{3}\text{Tr}(\mathbf{Q}^3) + \frac{C_t}{4}\text{Tr}(\mathbf{Q}^2)^2, \tag{12}$$

with $A_t$, $B_t$, and $C_t$ being material parameters determined by the phase transition temperatures [49]. Minimization of Eq. 12 gives us the equilibrium scalar order parameter:

$$S_{\text{eq}} = \frac{-B_t + \sqrt{B_t^2 - 24 A_t C_t}}{6 C_t}. \tag{13}$$

The elastic part, determined by the spatial derivatives of the tensorial order parameter, reads:

$$f_{\text{el}} = \frac{L_1}{2}\left(\frac{\partial Q_{ij}}{\partial x_k}\right)^2 + \frac{L_2}{2}\frac{\partial Q_{ij}}{\partial x_j}\frac{\partial Q_{ik}}{\partial x_k} + \frac{L_3}{2}\frac{\partial Q_{ij}}{\partial x_k}\frac{\partial Q_{ik}}{\partial x_j} + \frac{L_4}{2}\epsilon_{ijk}Q_{il}\frac{\partial Q_{kl}}{\partial x_j} + \frac{L_6}{2}Q_{ij}\frac{\partial Q_{kl}}{\partial x_i}\frac{\partial Q_{kl}}{\partial x_j}, \tag{14}$$

with **x** being the spatial coordinates and $\epsilon$ the antisymmetric tensor (Levi-Civita symbol). Summation over all indices is assumed in Eq. 14. These elastic distortion modes can be related to those in Frank-Oseen free energy (Eq. 3), as derived in Ref. [50] by neglecting the spatial gradient of the scalar order parameter and enforcing that $S=S_{\text{eq}}$:

$$\begin{aligned} L_1 &= \frac{2}{27 S_{\text{eq}}^2}(K_{33} - K_{11} + 3 K_{22}) \\ L_2 &= \frac{4}{9 S_{\text{eq}}^2}(K_{11} - K_{24}) \\ L_3 &= \frac{4}{9 S_{\text{eq}}^2}(K_{24} - K_{22}) \\ L_4 &= \frac{8}{9 S_{\text{eq}}^2} K_{22} q \\ L_6 &= \frac{4}{27 S_{\text{eq}}^2}(K_{33} - K_{11}) \end{aligned} \tag{15}$$

As shown in Eq. 15, the chirality of the material is characterized by the constant $L_4$ related to the twisting elasticity constant $K_{22}$ and the equilibrium scalar order parameter. Within one-constant approximation, the elastic moduli $L_2$, $L_3$, and $L_6$ vanish, and Eq. 14 reduces to an expression analogous to Eq. 7 [49]:

$$f_{\text{el}} \cong \frac{L}{2}\left(\frac{\partial Q_{ij}}{\partial x_k}\right)^2 + 2qL\epsilon_{ijk}Q_{il}\frac{\partial Q_{kl}}{\partial x_j}. \tag{16}$$

The surface anchoring free energy expression in terms of the tensorial order parameter, analogous to its counterpart in the director description in Eq. 9, adopts the harmonic potential form:

$$f_s = \frac{W}{2}\left(Q_{ij} - Q_{ij}^{\text{eq}}\right)^2, \tag{17}$$

with the equilibrium order parameter $\mathbf{Q}^{\text{eq}}$ constructed not only in terms of the easy axis, but from surface preferred value of $S$ as well. Other types of surface boundary conditions, planar or conically degenerate anchoring [51,52], for example, have also been developed to model various cases of surface confinements within the Q-tensor representation.

The diversity of descriptions of the energetic costs of perturbing CLC's orientational order reflects the large variety of practical situations that can arise and require such descriptions to capture the essential physics of phenomena involving CLCs. We will see below that this diversity of modelling CLC energetics is also mimicked by a large variety of different topological constructs in the order parameter field configurations that can arise within this soft matter system.

## 3. Homotopy theory of topological solitons and defects

CLCs exhibit particularly large variety of topologically nontrivial field configurations. These configurations include the ones of singular type (singular topological defects), containing regions of physical space where the order parameters cannot be defined, and the ones of nonsingular type (topological solitons), within which the structure of the λ(**r**) field is continuous everywhere, but it cannot be continuously morphed to a trivial, uniform state without destroying the order or introducing singular defects. Various topological structures can be classified based on mappings from the physical configuration spaces to the order parameter spaces (the manifolds of possible values of the order parameter) [8,18–22,24,26,48,53–55]. Typically, this is the mapping from spheres of various dimensions to the order parameter spaces that often are also multi-dimensional spheres. Therefore, the homotopy groups of spheres classifying these mappings are the most common (Fig. 2) and often utilized to label the different topologically distinct field configurations, though, as we will discuss below, there are also other homotopy group examples relevant to LCs and CLCs in particular. Algebraic topology describes how such spheres of various dimensions can wrap around each other, which is systematically characterized by the homotopy groups that describe the structure of topological spaces (without considering the precise geometry) [23,55,56]. In studies of topological solitons and singular defects, such classifications provide a means of summarizing topologically different structures in the order parameter fields, although the existence of a nontrivial element in the homotopy class does not guarantee their energetic stability or experimental observation [55]. The n-dimensional spheres (n-spheres, denoted as $\mathbb{S}^n$) are defined as sets of points equidistant from the origin in n+1 dimension, with an $\mathbb{S}^1$ circle being the 1-sphere embedded in 2D space ($\mathbb{R}^2$), $\mathbb{S}^2$ being an ordinary sphere embedded in 3D space ($\mathbb{R}^3$) and $\mathbb{S}^0$ being 0-sphere embedded in $\mathbb{R}^1$ that comprises 2 points equidistant from the origin in 1D, and so on [56–62]. The homotopy group labelled as $\pi_i(\mathbb{S}^n)$ is the i-th homotopy group that enlists the topologically different maps from $\mathbb{S}^i$ into $\mathbb{S}^n$, where none of the distinct mappings can be continuously deformed to the other mappings (Fig. 2) [22,53,55,61,62]. Algebraic topology results depend on the integers i relative to n, with $\pi_i(\mathbb{S}^n)=0$ for i<n (Fig. 2), which means that the corresponding homotopy groups are the trivial groups [61,62]. In the case of mappings between spheres of the same dimension (i=n), $\pi_n(\mathbb{S}^n)=\mathbb{Z}$, so that the spheres can be wrapped around spheres integer number of times for each map (Fig. 2). When i>n, a particularly interesting example of the mappings is called the Hopf fibration [61,62] (Fig. 2), which wraps $\mathbb{S}^3$ around $\mathbb{S}^2$ an integer number of times, $\pi_3(\mathbb{S}^2)=\mathbb{Z}$.

Since the singular defects in fields are discontinuities in the form of walls, lines and points, with the order parameter varying continuously outside these singular regions, one can surround them with spheres of the corresponding dimensions (say $\mathbb{S}^1$ for line defects and $\mathbb{S}^2$ for point singularities) and characterize how the field, like the vector or director field, varies around these spheres (Fig. 3) [22,53,55]. The order parameter spaces often also take the form of spheres. For example, the order parameter space for unit vectors in 3D space $\mathbb{R}^3$ is $\mathbb{S}^2$ (describing all possible orientations of the unit vector), but it becomes $\mathbb{S}^1$ when these unit vectors are forced to confine their orientations into a 2D plane $\mathbb{R}^2$ and becomes $\mathbb{S}^0$ when the unit vectors can only take orientations parallel or anti-parallel to the positive direction in $\mathbb{R}^1$. Therefore, the topologically distinct singular defects in unit vector fields can be classified with the help of maps from the i-spheres surrounding them to n-spheres describing their order parameter spaces [55]. Some of the simplest examples are illustrated in Fig. 3. Just like one can wrap one circle around the other an integer number of times (imagine wrapping a closed-loop rubber band around a finger), the structure mapped from $\mathbb{S}^1$ around a singular defect in 2D can wrap the $\mathbb{S}^1$ order parameter circle an integer number of times, $\pi_1(\mathbb{S}^1)=\mathbb{Z}$, indicating that singular defects with integer winding numbers exist in this system (Fig. 3a,b). Also, the structures of a vector field mapped from $\mathbb{S}^2$ around a singular point defect in 3D can wrap the $\mathbb{S}^2$ order parameter sphere an integer number of times (Fig. 3c), $\pi_2(\mathbb{S}^2)=\mathbb{Z}$, again defining the charges of all possible singular point defects in this system (Fig. 3) [21,22,53,55]. On the other hand, the fact that $\pi_1(\mathbb{S}^2)=0$ informs one that singular line defects in 3D unit vector fields are topologically unstable, so that they can be smoothly morphed to a uniform, topologically trivial state [55]. Likewise, since $\pi_2(\mathbb{S}^1)=0$, one cannot form topologically nontrivial point defects when the unit vectors are forced to take orientations confined to a 2D plane [22].

Nonsingular solitonic structures always have the field orientation well defined and, thus, may seem to be rather different from singular defects, but they can be classified based on the very same sphere-to-sphere maps (Fig. 2) [22,23,53,55,61,62]. In $\mathbb{R}^1$, a solitonic 360°-twist nonsingular wall in a unit vector field has the far field vector pointing upwards, and, thus, this configuration space can be "compactified" (by connecting the far-field regions of $\mathbb{R}^1$ with like-oriented unit vectors) into a circle, also referred to as one-sphere $\mathbb{S}^1$ (Fig. 3d,e) [39,60,63]. The topological class of the solitonic structures of this kind is then labelled by $\pi_1(\mathbb{S}^1)=\mathbb{Z}$, similar to the case of singular line defects for the 2D unit vector fields (Fig. 3a,b,d,e) [63]. The configuration space of solitonic topological structures embedded in the uniform far-field background in $\mathbb{R}^2$ can be compactified to $\mathbb{S}^2$ (e.g. by means of the stereographic projection), so that the nontrivial result $\pi_2(\mathbb{S}^2)=\mathbb{Z}$ from algebraic topology (in addition to classifying singular point defects like the one shown in Fig. 3c) also informs us that all possible topologically nontrivial structures in this case are characterized by the integer-valued 2D skyrmion numbers (Fig. 3f) [64,65]. Similarly, the configuration space in $\mathbb{R}^3$ with the uniform far-field is compactified to $\mathbb{S}^3$ through a higher-dimensional analogue of stereographic projection and the mathematical result from algebraic topology $\pi_3(\mathbb{S}^2)=\mathbb{Z}$ also implies that the Hopf indices of 3D spatially localized solitons in systems with order parameter spaces in the form of two-spheres also take integer values [60].

The multi-dimensional spheres cannot always represent the ground state manifolds for the order parameters [22,53,55,61]. Because of the non-polar nature of the LC director, representing all its orientations on $\mathbb{S}^2$ requires only half the sphere and leaves diametrically opposite points non-distinguishable from each other (Fig. 3g) [21,55]. The corresponding order parameter space is $\mathbb{S}^2/\mathbb{Z}_2 \equiv \mathbb{R}P^2$, a sphere with diametrically opposite points identified (Fig. 3g). One of the major differences as compared to the case of unit vectors is that $\pi_1(\mathbb{S}^2/\mathbb{Z}_2)=\mathbb{Z}_2$, meaning that singular vortex lines (disclinations) can be stable in 3D space of LCs [22,55], though only one type of such defect lines can be realized that is topologically different from the uniform state. These defect lines can have different local structures when embedded in 3D samples, including wedge disclinations with opposite signs of winding numbers (Fig. 3h,j) (which are topologically distinct when realized in 2D) and twist disclinations (Fig. 3i). In 3D, however, the defect line structures shown in Fig. 3h-j can be smoothly morphed one to another within $\mathbb{R}^3$ and are therefore topologically the same. In a similar way, unlike in the case of vector fields, one can realize nonsingular twist domain walls with only 180°-twist of the nonpolar director embedded in a uniform far field background (Fig. 3k,l), which are labelled by $\pi_1(\mathbb{S}^1/\mathbb{Z}_2) \equiv \pi_1(\mathbb{S}^1)=\mathbb{Z}$ [63].

The solitonic structures that exist in lower dimensions can be also embedded in higher dimensions while being translationally invariant with respect to some dimensions. For example, the $\pi_2(\mathbb{S}^2)=\mathbb{Z}$ solitons can be found as translationally invariant structures spanning $\mathbb{R}^3$ of LCs and magnets, either as individual spatially localized structures or periodic arrays [63]. When embedded in $\mathbb{R}^3$ in LC samples like glass cells of finite thickness, such solitons often terminate on $\pi_2(\mathbb{S}^2)=\mathbb{Z}$ point defects due to boundary conditions [63–66]. Similarly, translationally invariant solitonic walls $\pi_1(\mathbb{S}^1/\mathbb{Z}_2)=\mathbb{Z}$ are often embedded into finite-size structures in 2D by singular defects of the same class $\pi_1(\mathbb{S}^1/\mathbb{Z}_2)=\mathbb{Z}$; in 3D samples with all 3D orientations of director allowed, such twist walls are described by $\pi_1(\mathbb{S}^2/\mathbb{Z}_2)=\mathbb{Z}$ and can be embedded into a uniform background by the $\pi_1(\mathbb{S}^2/\mathbb{Z}_2)=\mathbb{Z}$ disclinations, forming one type of the so-called "cholesteric fingers" (a CLC finger of the 3rd type) [63]. The examples above illustrate a more general rule for imbedding lower-dimensional solitonic structures into a uniform background in higher dimensions with the singular defects of a homotopy class matching that of solitons [4,63]. The soft-matter topological solitons and defects have many topological counterparts in other branches of physics. For example, $\pi_3(\mathbb{S}^3)=\mathbb{Z}$ Skyrme solitons (Fig. 2) are used to model subatomic particles in high energy and nuclear physics [4,60,67], which is also the reason for often referring to their $\pi_2(\mathbb{S}^2)=\mathbb{Z}$ low-dimensional analogues in LCs and magnets as "baby skyrmions" [39,60,64].

Although exceptionally useful in classifying topologically distinct field configurations, homotopy theory does not provide the means for exploring the entirety of topological complexity of fields in soft matter even in cases when defects and solitons are embedded within a bulk of an ordered medium like the LC [66,68–71]. For example, a closed loop of a half-integer $\pi_1(\mathbb{S}^2/\mathbb{Z}_2)=\mathbb{Z}_2$ disclination is equivalent to a point defect $\pi_2(\mathbb{S}^2/\mathbb{Z}_2)=\mathbb{Z}$ in the far-field, but its hedgehog charge (topological charge of a point defect) depends on how this disclination is closed on itself, its local structure, twisting, knotting and possible linking with other defect loops [66,69–71]. Knowledge of this relation cannot be predicted solely by the homotopy theory but can be understood by invoking the analysis of the disclination's structure along the loop, its twist and writhe [69,70]. In other words, the homotopy theory identifies what the complex order parameter fields can be comprised of, but not how to obtain field configurations with desired $\pi_2(\mathbb{S}^2/\mathbb{Z}_2)$ hedgehog number by looping and knotting $\pi_1(\mathbb{S}^2/\mathbb{Z}_2)$ vortex lines or how to construct solitons with desired $\pi_3(\mathbb{S}^2/\mathbb{Z}_2)$ Hopf index by looping and knotting the 2D $\pi_2(\mathbb{S}^2/\mathbb{Z}_2)$ solitons. Moreover, when LCs interact with surfaces due to various boundary conditions, the topology of structures of these fields interplays with that of surfaces, which can be rather nontrivial and are a subject of ongoing studies [72–83].

The interpretation and classification of topological structures in CLCs depend on the level of description and the types of the order parameter used for this description. For nonpolar and vectorized director fields, the order parameter spaces are $\mathbb{S}^2/\mathbb{Z}_2$ and $\mathbb{S}^2$, respectively, with different types of possible topological defects and solitons described above, which will be discussed and illustrated within this review below. However, the use of all three director fields, $\lambda(\mathbf{r})$, $\chi(\mathbf{r})$ and $\tau(\mathbf{r})$, implies a more complete description of topological defects in CLCs, as we shall see in sections below. Throughout the sections below, we will emphasize how CLC defects and solitons can be described within different frameworks, similar to how free energetics of elastic distortions discussed in the previous sections could be theoretically described in different ways as well.

## 4. CLC disclinations and dislocations

Defect lines in CLCs are commonplace and include various disclinations and dislocations. For example, for vivid illustration purposes, they can be easily generated using the Grandjean-Cano wedge cells [26,29,84–86]. To fabricate such cells, a pair of mica or glass plates are placed at a small angle $\alpha$ to form a wedge cell such that the gap $d$ between the two confining surfaces varies along the bisector plane. Before filling-in with a cholesteric LC, the surfaces are often chemically treated to induce uniform planar anchoring, and to promote a uniform alignment of helical axes perpendicular to the substates. Fig. 4a, for example, shows a series of defect lines. The regions between the defect lines, called Grandjean zones, correspond to different numbers of helical layers $n \cong 2d/p$. A twist disclination separates $n=0$ and $n=1$ Grandjean zones (the first line in Fig. 4b). As $n$ increases starting from $n=1$, across the cell thickness gradient, the thin part of the wedge contains "thin lines" separated by distance $l=p/2\tan(\alpha)$ (Fig. 4a), while in regions with large cell gap one observes "thick lines" at the distances of $2l$. Also, the distance between the last thin and the first thick line is ~$1.5l$ (Fig. 4a). As originally theoretically explained by Kleman and Friedel [7], these defect lines are edge dislocations – a type of defect lines that breaks the translational symmetry of CLC quasilayers [20] – consistent with the change in the number of helical layers in the cross-sectional images of the Grandjean-Cano wedge (Fig. 4b). In the fluorescence confocal polarizing microscopy images [84,85,87,88] (Fig. 4b), each white/dark strip of the cross-sectional image represents a helical quasilayer with a rotation of the director by $\pi$. Accordingly, the Burgers vector **b** characterizing the Volterra process of these dislocations (associated with filling in or removing helical layers to become a trivial state) can be determined by the corresponding translational displacement [9,20,26]. For instance, a thin line separating the regions of $n=2$ and $n=3$ is represented by **b**||z with magnitude $b=p/2$; a thick line marking the boundary of regions $n=23$ and $n=25$ has Burgers vector of magnitude $b=p$ and oriented parallel to the z direction [84] (Fig. 4b).

In Grandjean-Cano wedges with a strong in-plane surface anchoring, both $b=p/2$ and $b=p$ dislocations are located in the bulk of the cell, at or close to the bisector plane. The edge dislocation fragments at different sample depths are inter-connected by energetically metastable structures called kinks (Fig. 4c) [26,84,85]. Viewed along the dislocation line, the molecular director rotated 90° around the helical axis (dark ↔ white in the image) when the z coordinate is changed by $p/4$; a second kink in the same direction restore the structure by another 90° rotation of $\lambda$ and another vertical displacement of $p/4$ (Fig. 4d). The two structures of $b=p/2$ dislocation, which will be discussed in detail in Section 5, have distinct energetic properties [3]. Therefore, one can often see kinks traveling along defect lines in a freshly prepared CLC cell, showing structural and positional relaxation of the dislocation positions, often slowly evolving with time (days).

Other than dislocations that mark the changing cell gap, thick defect lines are often observed perpendicular to the equilibrium dislocations and parallel to the thickness gradient of the wedge (Fig. 5a-c). These defect lines have been shown to be pairs of dislocations with opposite signs of the Burgers vector, adding up to net zero $b=0$ (Fig. 5d). In the case of two parallel symmetric $b=p$ dislocations with the same depth location (z coordinate), the simplest such energetically metastable structure is referred to as a "Lehmann cluster" [89] (Fig. 5e). Since these assemblies of dislocations are topologically trivial from the standpoint of the quasilayer structure, they shrink quickly to an unperturbed helical state under external disturbance. Defect lines of Burgers vector $b=0$ can be found connecting $b=p/2$ dislocations (Fig. 5a), $b=p$ dislocations (Fig. 5b) or even one of each (Fig. 5c). Near the joint of these defect lines, tilt and shift of dislocations to a different depth z level are observed. The topology, however, is preserved within the positional deviation. Moreover, the sum of line tensions of individual dislocations should be zero under mechanical equilibrium. The different angles $\phi$ made by thin and thick lines with $b=0$ dislocation, therefore, inform us that dislocations with $b=p/2$ are characterized by larger values of line tension and higher cost of elastic energy than that of $b=p$ type. This experimental observation can be explained by the energetics associated with the split core structures of the defects, as we discuss in the following section. The lamellar-like nature of CLCs in different geometries was used to test nonlinear theory of smectic elasticity by probing layer profiles around edge dislocations in different geometries, showing good agreement between experiments and theoretical models even before similar experiments could be done for the actual lamellar LCs [85,90,91]. The models describing CLC dislocations and disclinations in this geometry were introduced by Kleman and Friedel in 1969, even though it was several decades later that researchers could directly visualize such defects by means of direct 3D imaging of the director field [84].

## 5. Quaternion representation of disclinations in chiral nematics

As introduced previously, structures of CLCs can be most fully described by the orthonormal frame $\{\lambda(\mathbf{r}), \chi(\mathbf{r}), \tau(\mathbf{r})\}$ with three unoriented, nonpolar director fields (Fig. 6a). The corresponding order parameter space of the CLC yields a different homotopy group representation of field configurations. Due to the ensuing orthorhombic point group symmetry $D_2$, the topology of a CLC within this representation is hence identical to biaxial nematic material with brick-shaped ($D_2$) building blocks [55] (Fig. 6b). With the nonpolar directors free to rotate in 3D space ($\mathbb{R}^3$), the order parameter space of a CLC can be expressed as $O(3)/D_{2h}$ or $SO(3)/D_2$ [21,92]. Thus, the classification of singular defect lines in a CLC is characterized by the first homotopy group $\pi_1(SO(3)/D_2)=Q_8$, the quaternion group [21,23,24,26,55,92–97] (with the multiplication table provided in Fig. 6c). There are exactly eight topologically different defect lines that could possibly form in such CLC systems. Within the eight group elements of the quaternion group $Q_8$, six elements ($\pm\lambda$, $\pm\chi$, $\pm\tau$) correspond to defect lines with half-integer winding numbers, either +1/2 or -1/2 (see examples in Fig. 6d-f), the element "1" corresponds to the trivial state with no singularity in any of the director fields (Fig. 1b), and the element "-1" represents disclinations with a winding number of one [23,55,98]. Moreover, for all nontrivial structures of cholesteric disclinations, two out of the three director fields exhibit singularities where the orientation of directors cannot be defined, while the third director is continuous throughout the structure. This property of "semi-defects" suggests a simple terminology for half-integer disclinations: a X line is a cholesteric disclination nonsingular only in X field [18,20,23]. For example, $\chi$ lines are disclinations with uniform helical axis field, while singularities can be found in $\lambda(\mathbf{r})$ and $\boldsymbol{\tau}(\mathbf{r})$ fields (Fig. 6d) and the structure is known as a screw dislocation [18,23,48]. This classification of line defects in CLCs as $\lambda$, $\tau$ and $\chi$ lines was originally proposed by Kleman and Friedel and is currently widely used [7]. Interestingly, the director alignments of two different $\lambda$ lines, one with defects in $\chi(\mathbf{r})$ and $\boldsymbol{\tau}(\mathbf{r})$ fields having a winding number +1/2 ($\lambda^{+1/2}$, Fig. 6e) and the other having a -1/2 defect in each of $\chi(\mathbf{r})$ and $\boldsymbol{\tau}(\mathbf{r})$ fields ($\lambda^{-1/2}$, Fig. 6f), can be used to illustrate $\tau$ lines with the very same $\chi(\mathbf{r})$ directors while swapping $\lambda$ and $\tau$ field configurations (one then obtains defect lines with singular defects in $\lambda(\mathbf{r})$ and $\chi(\mathbf{r})$ but not in $\boldsymbol{\tau}(\mathbf{r})$, hence $\tau$ lines). The idea of interchangeability of $\lambda(\mathbf{r})$ and $\boldsymbol{\tau}(\mathbf{r})$ fields (with the same $\chi(\mathbf{r})$ alignment) is consistent with the Lubensky-de Gennes coarse-grained model of CLC elasticity (Eq. 3), in which deformations are described only in terms of $\chi(\mathbf{r})$. Topologically, in some sense, it also suggests an explanation of why most of the dislocations are found to be comprising combinations of $\lambda$ and $\tau$ lines in their cores, which will be discussed below.

In accordance with the conventional classification of disclinations using five conjugacy classes (1, -1, $C_\lambda$, $C_\chi$, $C_\tau$) [23,55] (as distinct colors in Fig. 6c), the name "$\lambda$ line" itself does not distinguish the quaternion group elements "$\lambda$" and "-$\lambda$" [18,23,55] whereas they have different winding numbers, which can be found from the cross-section imaging of the disclinations [84,99]. The detailed discussion of apparent ambiguity [23,55] in the one-to-one assignment of the $Q_8$ group elements to disclinations in CLCs is beyond the scope of this review, and for the following paragraphs we use superscript to indicate only the winding number.

Following the notation defined previously, all defect lines in CLCs can be interpreted as $\lambda^{\pm 1/2}$, $\chi^{\pm 1/2}$, $\tau^{\pm 1/2}$, and -1 disclinations or their combination, as in the cores of dislocations. The simplest way to identify the quaternion representation of a line defect is to find out which director field is nonsingular along the structure, and to obtain the winding number from the director alignments of the other two fields. The cross-section of an edge dislocation of Burgers vector $b=p/2$, as an example, can have two distinct configurations of director alignments [84] (Fig. 7a,b). One of the thin line structures within the dislocation core is a $\tau$ line with winding number -1/2 on the side with lower number of helical layers and a $\lambda^{+1/2}$ at the opposite side, and is subsequently labelled with $\tau^{-1/2}\lambda^{+1/2}$ (Fig. 7a). The other structure, denoted as $\lambda^{-1/2}\tau^{+1/2}$, has the disclinations distributed distinctly with the same winding numbers (Fig. 7b). The dislocations with these core structures have identical $\chi(\mathbf{r})$ field alignment, within which one can find a pair of half-integer defects with the opposite winding numbers ($\pm 1/2$), and remind us of the topological similarity between $\lambda(\mathbf{r})$ and $\tau(\mathbf{r})$ fields. By swapping $\lambda(\mathbf{r})$ and $\tau(\mathbf{r})$ field arrangement of $\tau^{-1/2} \lambda^{+1/2}$, one gets to the director structures of $\lambda^{-1/2} \tau^{+1/2}$ (Fig. 7a,b). Furthermore, the associated elastic energy cost is higher for $\lambda^{-1/2} \tau^{+1/2}$ compared with $\tau^{-1/2} \lambda^{+1/2}$ [3,18,84], which can be qualitatively explained by the larger relative volume occupied by the disclination with positive winding number and the stronger elastic distortion within the $\tau$ lines.

As for edge dislocation of Burgers vector $b=p$, similar analysis of the composing directors shows that two parallel $\lambda$ lines with the opposite winding numbers contribute to the jump in the number of helical layers (Fig. 7c). Again, the positions and winding numbers of the two disclinations $\lambda^{-1/2} \lambda^{+1/2}$ are revealed by the two half-integer defects in the $\chi(\mathbf{r})$ field of the cross-section image (Fig. 7c), which resemble that of $b=p/2$ dislocations. Though lacking stability on the energy landscape, $\tau^{-1/2} \tau^{+1/2}$ is another hypothetically possible structure of a $b=p$ dislocation but with director singularities in $\lambda(\mathbf{r})$ instead of $\tau(\mathbf{r})$ field [84]. One can picture this energetically unstable dislocation core with $\tau^{-1/2} \tau^{+1/2}$ disclinations by exchanging the two director fields of a $\lambda^{-1/2} \lambda^{+1/2}$ type (Fig. 7c), analogous to the case of thin lines discussed above. Furthermore, +1/2 defect in $\chi$ field is found at the side with a larger number of helical layers since the elastic distortion of $\chi(\mathbf{r})$ energetically favors such arrangement in an equilibrium structure of dislocation within a uniform $\chi$ far-field. The +1/2 defect line within a dislocation core, therefore, is always located at the thicker side of the liquid crystal wedge cell, which has also been revealed experimentally through imaging with fluorescence confocal polarizing microscopy [84,85].

Another common line defect in CLCs is the so-called "Lehmann cluster" (Fig. 5e), which can be decomposed into four lambda lines, seen by visualizing the three director fields (Fig. 7d). As shown by the $\chi(\mathbf{r})$, two of these λ lines have +1/2 winding number and the other two have the winding number of -1/2 (Fig. 7d). With a director organization similar to the combination of two $\lambda^{-1/2} \lambda^{+1/2}$ dislocations in the opposite direction, Lehmann cluster is energetically metastable and has zero Burgers vector. However, it terminates on point defects and its cross-section can be viewed as a full 2D skyrmion [100], as we will discuss further below. Its **λ-τ**-exchanged counterpart with four τ lines is energetically unstable and not observed due to high energetic costs.

Besides dislocations composed of disclinations, the elastic energy of individual disclinations can be numerically computed using tensorial order parameter given by Eq. 14. (In the cases of half-integer defects, directors cannot be vectorized without introducing additional singularities like wall defects connecting them.) Under parameters for common LCs like 5CB, τ lines have higher elastic energies compared with χ lines, with λ lines being most stable disclinations due to the absence of singularity in the molecular field. Furthermore, since these disclinations are represented by different elements in the same homotopy group (the quaternion group $Q_8$), their interaction with and transformation into each other are thoroughly characterized by the multiplication rules of the group [23,54,55,96,101,102] (Fig. 6c). This provides an explanation for the lack of experimental observation of an isolated τ line, which is expected to quickly transform into pairs of other disclinations due to its high energetic costs. However, as far as we are aware, there have not been clear experimental demonstrations of these multiplication rules involving cholesteric disclinations beyond the representation of defects by homotopy group elements. By employing director analysis of dislocations and disclinations in CLCs, we can see that the different types of dislocation cores in CLCs effectively comprise distinct arrangements of λ and τ lines embedded within helical quasilayers, or equivalently, pairs of half-integer defects within uniform far-field in terms of **χ(r)** directors. The quaternion interpretation also provides deep insights into the geometries and energetics of the CLC defect lines. Interestingly, the classification of CLC disclinations as λ, τ and χ lines was introduced by Kleman and Friedel [7] well before the quaternion interpretation and homotopy (also co-invented by Kleman) theory classification of CLC defects was introduced and appreciated [21,22,55,92], but it naturally became an important part of it.

## 6. Fractional, full and multi-integer skyrmions

CLCs host a large variety of topological solitons. To understand them, in principle, it is sufficient to characterize field configurations only using the material field, **n(r)**≡**λ(r)**. Two-dimensional, translationally invariant skyrmions and merons that exist in chiral magnetic spin textures can be also found in CLCs and vice versa, due to the similarities between the two systems in terms of free energy and accessible degrees of freedom. For example, a variety of magnetic "baby skyrmions", particle-like low-dimensional analogues of Skyrme solitons originally introduced in particle physics [60], have been realized in solid-state magnetic systems [103]; it is instructive to overview them here before introducing their LC counterparts. With unit magnetization order parameter of chiral magnets having the order parameter space $\mathbb{S}^2$, these smooth vector configurations in two-dimensional space (which can have embodiments of translationally invariant structures in the three-dimensional space) are characterized by the homotopy group $\pi_2(\mathbb{S}^2)=\mathbb{Z}$. This result implies that all structures of these skyrmions correspond to a localized spatial distribution of unit vectors wrapping around the order parameter space $\mathbb{S}^2$ integer number of times (Fig. 8). Thus, 2D magnetic skyrmions are assigned with integer topological numbers $N_{sk}$, sometimes also called vorticity (Fig. 8) [4,103,104]. The topological degree $N_{sk}$ of a baby skyrmion is also referred to as the skyrmion number, which counts the number of times the field configuration of the magnetization unit vector field $\mathbf{m}(\mathbf{r})$ wraps around its target space $\mathbb{S}^2$ [63,104,105]:

$$N_{sk} = \frac{1}{4\pi} \iint dxdy\, \mathbf{m} \cdot (\partial_x \mathbf{m} \times \partial_y \mathbf{m}). \tag{18}$$

In addition to $N_{sk}$, the helicity parameter $\gamma$ is used to describe their geometry and may determine their stability in different material systems. The parameter $\gamma$, in principle, can take arbitrary values, though the corresponding structures with the same $N_{sk}$ but different $\gamma$ are able to be smoothly inter-transformed into each other [103,104]. When $\gamma=0$ or $\pi$, linear Neel domain walls can be found extending from the center of the skyrmion to the periphery, and the structures are thus referred to as Neel-type skyrmions [106]. Bloch type skyrmions with $\gamma=\pm\pi/2$, in contrast, comprise rotating spins orthogonal to the radial direction and are found being stabilized in chiral magnetic systems (Fig. 8). The skyrmions with $N_{sk}=1$ and $\gamma=\pm\pi/2$ correspond to the lowest energy and are found in the so-called A-phase of chiral magnetic systems (we note that the sign of $N_{sk}$ depends on the convention; with a different convention, the same skyrmion can be assigned $N_{sk}=-1$ [107]). For our $N_{sk}$ sign-defining convention used in Fig. 8, the $N_{sk}=-1$ skyrmions are energetically much more costly in chiral magnets because of containing regions with opposite handedness of twist within them. Both Neel-type and Bloch-type skyrmions can be smoothly embedded in the uniform far-field magnetization background as individual localized objects. The color scheme of unit vector orientations adopted in Fig. 8 shows how localized structures of the 2D skyrmions contain all possible values (orientations) of the order parameter **m(r)**.

The nonpolar counterparts of the 2D magnetic skyrmions, the LC skyrmions (Fig. 9), have also been extensively studied analytically, numerically and experimentally (Fig. 10) [4,63,107–112]. Represented by the homotopy group $\pi_2(\mathbb{S}^2/\mathbb{Z}_2)=\mathbb{Z}$, LC skyrmions can be illustrated in similar ways as in magnets. That is, the skyrmion number $N_{sk}$ labels the topological charges and the helicity $\gamma$ reflects the chirality of the LC material, but with each elementary skyrmion wrapping around the order parameter space ($\mathbb{S}^2/\mathbb{Z}_2$) exactly twice due to the nonpolar nature of the host medium (Fig. 9). Upon smoothly vectorizing the nonpolar LC director field within the entire two-dimensional structure of the translationally invariant LC skyrmion's cross-section, the order parameter space becomes $\mathbb{S}^2$ and the LC skyrmion's texture becomes analogous to that of the skyrmions in magnets, with $N_{sk}$ then again determined by Eq. 18. An important difference from magnetic skyrmions, however, is that this "decoration" of the nonpolar LC director with unit vectors can be done while aligning the far-field vector along two anti-parallel directions of the far-field director, which yields opposite signs of the skyrmion numbers, reflecting the nonpolar nature of the LC host medium. Therefore, the topological invariant $N_{sk}$ of an isolated 2D LC skyrmion can be defined only up to the sign, a consequence of the nonpolar nature of the director field. This property of LC skyrmions resembles that of 3D point defects in nematic LCs, described by the same homotopy group, $\pi_2(\mathbb{S}^2/\mathbb{Z}_2)=\mathbb{Z}$ [79], where, unlike for point defects in vector fields, the hedgehog charge is also defined up to the sign. Like for chiral magnets, CLC skyrmions with twisted structures matching that of the chiral host medium are energetically the most stable as they help realizing the CLC's tendency to twist. Because of the common LC's strong interactions with confining surfaces, realization of skyrmions as both individual objects and arrays requires rather soft but controlled surface boundary conditions [63,108,113]. Figure 10 shows examples of CLC skyrmions, both observed experimentally and modelled numerically for such conditions [63,108,113].

When embedded in 3D samples, the CLC and magnetic 2D analogues of Skyrme solitons have topologically protected translationally invariant 2D tube-like structure that cannot be eliminated from a uniformly oriented background without destroying the order or introducing singular defects. Differently, tubes of merons, also known as fractional skyrmions (Figs. 11,12), are typically accompanied by singular line defects (for CLCs) or by other merons with the same-sign or opposite-sign fractional charges, especially when embedded in the uniform far-field background [111,114]. Spins of chiral magnets or rod-like molecules of CLCs within the simplest type of a fractional skyrmion (double twist tube) contained within blue phases are parallel to its axis at the center, twisting radially outwards (Figs. 11,12), with the configurations identical to the central part of full skyrmions (Figs. 8,9). For example, blue phases of CLCs can be interpreted as various arrays of double-twist tubes that are fractional skyrmions (merons or half- and quarter-integer skyrmions) [40,107–109,111,115–120], including their cubic and hexagonal lattices, where fractional skyrmions are accompanied by singular line defects. As compared to the solid-state magnets, the possible structures involving the skyrmionic field configurations are enriched by the nonpolar nature of chiral LCs, which can also host topologically stable singular line defects [40,41,109,111,118,119,121–130].

The study of integer-strength disclinations in the bulk of LCs, including CLCs, concluded that they either split into half-integer lines or "escape in the third dimension" upon being embedded in 3D and being allowed 3D orientations of the order parameter [9,26,55,131–137]. Interestingly, the director structures within merons closely resemble escaped integer-strength disclinations in the bulk of LCs (Fig. 12), with the topological invariant $N_{sk}$ of half-skyrmions inheriting half of the pre-escape disclination's original winding number. Though singular defect lines are no longer stable once 3D orientations of director are allowed as $\pi_1(\mathbb{S}^2/\mathbb{Z}_2)=0$, nontrivial field configuration become merons (half skyrmions in $\mathbb{S}^2/\mathbb{Z}_2$), inside the 2D cross-sectional plane of the LC bulk. Therefore, the fractional skyrmions, with the molecular alignment field being fully nonsingular, can be interpreted in the historic LC language as 3D-escaped disclinations with pre-escape winding numbers of $2N_{sk}$ in a material with different properties related to $\gamma$. For example, the half-skyrmions with a radial structure of the director in their periphery are found in thick capillaries with the so-called "escape of the director in the 3$^{rd}$ dimension" [133–136]. Within Schlieren textures of thick nematic slabs confined between plates with tangentially degenerate boundary conditions, practically all configurations of half-skyrmions shown in Fig. 12 can be found existing in the bulk of LC cells, albeit terminating on various boojum surface point defects at the confining surfaces (hypothetically, such boojums can be avoided when surface boundary conditions are sufficiently soft) [133–136]. The "thick" lines (threads) that gave the Greek-origin name "nematic" to the most common type of LCs, which are 3D-escaped integer-strength lines [23], also can be interpreted as merons, similar to objects that recently fascinate researchers and play important roles in solid-state magnetic systems [114]. Moreover, Bloch-type merons ($\gamma=\pm\pi/2$) are naturally found to be stable in CLCs [40,63,108,111] and recent studies revealed fascinating thermodynamically stable 2D crystals of such half-skyrmions in thin CLC cells [109,111]. Under specific confinement conditions, the stability of such half-skyrmion lattices (also containing singular defect arrays) in chiral LCs does not require applying external fields [109,111], which is different from the chiral magnets.

The common CLC fingers of the first and second types can be interpreted as being composed of the nonsingular λ-disclinations, or fractional skyrmions (Fig. 12b,c). The structures of these cholesteric fingers show how the fractional values of $N_{sk}$ add to unity in the finger of the second type (comprising one half skyrmion and two quarter skyrmions) and to zero for the finger of first type (four quarter-skyrmions), in both cases being embedded to the uniform far field background (Fig. 12b,c). The mapping to the vectorized director field to the order parameter space $\mathbb{S}^2$ from a cholesteric finger of the second type (cross-section shown in Fig. 11b) indeed gives an integer skyrmion number $N_{sk}=1$. Therefore, one can morph an axisymmetric translationally-invariant skyrmion tube into a cholesteric finger of the 2$^{nd}$ type by rotating individual vectors by 90°. It has been shown recently that such morphing can take place in CLCs with negative dielectric anisotropy upon application of electric field and actually can lead to squirming motion of such solitons when this field is periodically modulated [64]. Another example of a 2D skyrmion that can be decomposed into fractional skyrmions (λ-disclinations) is the skyrmion embedded in a helical structure of CLCs or chiral magnets, which is the already familiar Lehman cluster shown in Fig. 5e. Such Lehman clusters (skyrmions) have been known to the CLC community for many years, though their 2D skyrmion nature was not fully appreciated until recent interest that similar structures attracted in chiral magnets [100,126]. Just like the skyrmions embedded in a uniform far-field background, their counterparts in the helical field background can terminate or nucleate on point defects of an integer hedgehog charge matching $N_{sk}$. The four λ-disclinations within such skyrmions can be each assigned fractional $N_{sk}$ values, with all four fractional charges adding to unity and then being embedded in the uniformly twisted quasi-layered structure of the CLC. Equivalently, a $\lambda^{-1/2} \lambda^{+1/2}$ type $b=p$ dislocation (half of a Lehmann cluster) can be associated with a meron (half-skyrmion), as also noted in Ref. [138]. In thin CLC films, individual fractional skyrmions could be also embedded into a uniform background with the help of singular disclination lines, as well as 2D crystals comprising half-skyrmions and singular line defects could form [109,111]. Stabilized by the thin-cell confinement at no external fields, such 2D lattices of CLC half-skyrmions and singular defects cannot have magnetic analogues because they can exist only in nonpolar LCs where half-integer singular defect lines are allowed [109,111].

Translationally invariant tube-like structure of CLC Bloch skyrmions can be also described using the quaternion representation for disclinations in CLCs (though $\chi(\mathbf{r})$ and $\tau(\mathbf{r})$ fields are ill-defined at peripheries where $\lambda(\mathbf{r})$ is nonchiral). With integer winding numbers, skyrmions in chiral LCs corresponding to the cross-section of -1 lines in the terminology of the quaternion representation. As an instance, the $\chi(\mathbf{r})$ directors in a $N_{sk}=1$, $\gamma=\pm\pi/2$ skyrmion are aligned along the radial direction, forming a point defect with +1 winding number on the two-dimensional plane (Figs. 9,12). It is worth noting that the -1 disclination can be morphed into several different configurations with $\lambda(\mathbf{r})$ field singular or nonsingular while preserving the continuity of director fields around the defect line, and that $\gamma=\pm\pi/2$ skyrmions are the cross-section of -1 lines with least elastic free energy (calculated using Frank elasticities Eq. 2, for example). With the introduction of $\chi(\mathbf{r})$ and $\tau(\mathbf{r})$ field characterizations for chiral systems, 2D and even 3D topological solitons in CLCs, conventionally described as nontrivial structures with a continuous $\lambda(\mathbf{r})$ field, can now be interpreted as singular defects of $\chi(\mathbf{r})$ and $\tau(\mathbf{r})$ fields, as we shall see discussed in sections below.

Skyrmions in magnets drive much excitement because of their potential for applications in spintronics, including data storage [41,121–123,127–130]. Density of such topologically-encoded information could be increased by using skyrmions with varying topological degrees (whose distinction is topologically protected) [107]. CLCs recently provided insights into how high-degree skyrmionics structures can form [107] as stable chiral composite skyrmion bags. To realize them, one places multiple single antiskyrmions (each with elementary skyrmion number topological invariant) next to each other within a stretched elementary skyrmion , thus forming the skyrmion bags (Fig. 13) [107]. Moreover, multiple nested structures can be formed, with, say, antiskyrmion bags within skyrmion bags and skyrmions within them, and so on [107]. This yields nonsingular skyrmionic structures with arbitrary degrees and of both positive and negative signs because this design allows for wrapping and unwrapping $\mathbb{S}^2$ by mapping $\lambda(\mathbf{r})$ from the sample's 2D plane by controlled numbers of times in a non-alternating fashion [107]. The total degree of a skyrmion bag with $N_A$ antiskyrmions is $N_A-1$. More complex structures with antiskyrmion bags inside skyrmion bags have a net degree $N_A-N_S$, where $N_S$ is the total number of skyrmions; counting $N_S$ and $N_A$ also includes the nested skyrmion and antiskyrmion bags. While we noted above that $N_{sk}$ of CLC skyrmions can be defined up to the sign due to LC's nonpolar nature, it is important to keep track of the relative signs of skyrmions within composite skyrmionic structures, like the skyrmion bags. A way to do this involves vectorization of the director field, an approach also used to characterize 3D textures in nematic LCs with multiple hedgehog point defects, where hedgehog charges of these defects within 3D textures could be analyzed by smoothly vectorizing the director field [79].

The stability of skyrmions and skyrmion bags in CLCs require careful selection of experimental conditions and materials [63,107], where important roles are played by soft but well-defined perpendicular boundary conditions on confining surfaces, elastic anisotropy, confinement, applied fields, etc. The similarities of Hamiltonians and soliton topologies between CLCs and chiral magnets allows for using CLCs as model systems to provide insights into solitonic structures that can be also realized in chiral magnets [40,64,107,113].

While the recent interest in the 2D skyrmion structures of CLCs was re-ignited by the very active area of research in chiral magnets and other branches of condensed matter and particle physics [41,67], such non-singular structures have a long history of studies by the LC community. Furthermore, CLCs offer unique experimental accessibility of such structures, their facile control by external fields and a richer range of possibilities enabled by allowed half-integer defect lines that can co-exist with full and fractional skyrmions or compete for energetic stability under different conditions [109,111]. While surface anchoring boundary conditions were shown to be the key enabling realization of fractional, full and multi-integer skyrmions as translationally invariant topological structures [63,65,113], what we will discuss next are CLC torons that arise as stable configurations under different boundary conditions.

## 7. Torons and twistions with fractional or full skyrmions within them

In addition to skyrmions and skyrmion bags [63,107,108], confined CLCs can also host structures called "torons" with both skyrmion-like and Hopf fibrations features when a chiral LC is confined by substrates treated for perpendicular alignment boundary conditions of the director [65,120,139]. When the separation gap $d$ of the confining planes is approximately equal to the CLC's pitch $p$, the tendency to twist is incompatible with the strong perpendicular boundary conditions that induce the background of unwound far-field director $\lambda_0$ [28,63,65,139]. The solitonic configurations that emerge incorporate energetically-favorable localized twist while meeting boundary conditions [139,140]. The nonpolar and vectorized configurations of the simplest toron is shown in Fig. 14 [63]. In the cell midplane between confining substrates, the elementary toron features a $\pi$-twist of $\lambda(\mathbf{r})$ radially from the center in all directions (Fig. 14a-c) and smoothly meets the uniform far-field $\lambda_0$-periphery [63]. Vectorized director alignment from the toron's midplane cross-section (Fig. 14c) maps to fully cover $\mathbb{S}^2$ once (inset of Fig. 14d), like for an elementary skyrmion. This skyrmion tube, however, terminates at point defects that are enforced by the uniform boundary conditions at surfaces (Fig. 14b-d). Both top and bottom point defects are self-compensating hedgehogs of opposite charge in the vectorized $\lambda$ field and, like elementary skyrmions, are labelled by $\pi_2(\mathbb{S}^2)=\mathbb{Z}$ (or $\pi_2(\mathbb{S}^2/\mathbb{Z}_2)=\mathbb{Z}$ for the nonpolar case) [55,63]. The structure of torons can be characterized further by probing streamlines tangent to $\lambda$ (Fig. 14), which form various torus knots, resembling the ones found in toroidal DNA drops [141,142]. Regions near the toron's circular axis resemble fragments of $\mathbb{S}^3$ to $\mathbb{R}^3$ stereographic projection [65]. Differently from biopolymer drops, toron's $\lambda$-twist rate changes smoothly as one moves away from its axis, accommodating effects of confinement and presence of the singular defects (Fig. 14e). Incompatible with Euclidian 3D space [141], 3D twist is inherently frustrated, but the geometry of fiber bundles shows how LC embeds it into toron's volume [141]. Torus knots can be identified within such structures with a series of streamlines of $\lambda(\mathbf{r})$. While the elementary toron is a skyrmion tube terminated on point defects, it also has an interpretation inspired by the torus-knot-like streamlines tangent to $\lambda(\mathbf{r})$. One can also think of it as a half-skyrmion double-twist tube (or a tube of twist-escaped integer-strength line) forming a circular loop and compensated near substrates by a pair of hyperbolic point defects [139]. It has been shown that the point defects within these torons can open up into singular half-integer defect loops [139]. Therefore, within the 2D axially symmetric cross-section of a toron with two such singular loops [139], the half-skyrmion (meron and also the twist-escaped integer-strength line) is compensated by two singular defect lines in a way similar to what has also been shown for individual linear half-skyrmions embedded into a uniform background and periodic lattices [111,120]. Torons have been generated by laser tweezers and temperature quenching from isotropic phase, both as individual objects and in periodic arrays [115,143,144], with and without lattice defects. Toron lattices have been used as diffractive optical elements whereas lattices with edge dislocations could be utilized as generators of optical laser vortices [117]. In addition to CLCs, torons have been recently discovered in chiral solid-state magnetic systems, where magnetic torons adopted the same name as in CLCs [145].

In addition to the elementary torons with $\pi$-twist of $\boldsymbol{\lambda}(\mathbf{r})$ from their central axes to the periphery in all radial directions, torons with larger amounts of such twist also exist. For example, such more complex torons can exhibit $3\pi$, $5\pi$ and larger amounts of twist within axisymmetric toron structures, as found in recent experiments, though we will not discuss them in detail here [110]. Solitonic topological structures and singular point defects also co-exist within hybrid structures called "twistions", configurations that embed spatially localized twisted regions into a uniform far-field $\boldsymbol{\lambda}_0$-background but lack axial symmetry and (unlike torons) contain more than two point defects [146]. Within their structure, $\boldsymbol{\lambda}(\mathbf{r})$ typically twists from the interior to periphery by $\pi/2-\pi$, though twistions with larger amounts of such twist exist too, analogously to what was discussed above for torons [110,146,147]. For example, the simplest twistions contain stretched loops of merons and four self-compensating hyperbolic point defects. Similar multi-point-defect configurations with solitonic λ(**r**) in-between have been also reported for CLC drops [128,148].

## 8. 3D topological solitons

### 8.1. Hopfions

The fully three-dimensional topological Hopf soliton, also called "hopfion", was recently observed experimentally and modelled numerically [39]. This topological soliton is a physical embodiment of the mathematical Hopf fibration's topology in the unit vector and director order parameters [4,39]. Although hopfions have been studied in both nonpolar CLCs and in complex fluids formed by colloidal dispersions of magnetically monodomain platelets within a CLC, referred to as chiral ferromagnetic LCs [4], here we will use the latter to introduce topological properties of hopfions. The structure and topology of hopfions can be effectively described with the concept of "preimage", the inverse mapping from a single point on $\mathbb{S}^2$ to the CLC ferromagnet's 3D physical space with the same unit magnetization field **m**(**r**) orientation (Fig. 15a). For hopfions, preimages of all $\mathbb{S}^2$-points are closed loops [60] (Fig. 15a,b). Imbedded into a uniform **m**$_0$ and localized in all three spatial dimensions (Fig. 16), hopfions are classified on the basis of maps from $\mathbb{R}^3 \cup \{\infty\} \cong \mathbb{S}^3$ to the ground state manifold $\mathbb{S}^2$ of 3D unit vectors, $\pi_3(\mathbb{S}^2) = \mathbb{Z}$ [23,55,60]. Topologically distinct hopfions are characterized by the Hopf index $Q \in \mathbb{Z}$ with a geometric interpretation of the linking number of any two closed-loop preimages (Fig. 15c,d). Most of the three-dimensional sample ($\mathbb{R}^3$ space) is occupied by the preimage of the north-pole in $\mathbb{S}^2$ corresponding to **m**$_0$ (Fig. 16) [39], except for the interior of a torus-embedded region, within which all other preimages are smoothly packed. Preimages with the same polar angles but different azimuthal angles tile into tori; then, tori corresponding to different polar angles nest within each other, all imbedded within the biggest torus that has all the preimages in its interior, except for the $\mathbf{m}_0$-preimage that is in its exterior (Fig. 16) [39]. In addition to being derived from its geometric interpretation as the preimage linking number, $Q$ can be computed explicitly by integrating a topological charge density in either $\mathbb{S}^3$ or $\mathbb{R}^3$. For a solitonic unit vector field $\mathbf{m}(\mathbf{r})$ in $\mathbb{R}^3$ with a uniform far-field $\mathbf{m}_0$ [32]:

$$Q = \frac{1}{64\pi^2} \int dV \, \epsilon_{ijk} \, A_i \, B_{jk}, \tag{19}$$

where $B_{ij} = \epsilon_{abc} m_a \, \partial_i m_b \, \partial_j m_c$, $A_i$ is defined as $B_{ij} = \frac{1}{2}(\partial_i A_j - \partial_j A_i)$. Stable hopfions in physical systems ranging from elementary particles to cosmology have been predicted by Faddeev, Niemi, Sutcliffe and many others [60,149,150], as well as demonstrated experimentally as stable solitons in colloidal ferromagnets and LCs [39,110,147]. Nonlinear optical 3D imaging was utilized to unambiguously identify topological solitons, revealing an experimental equivalent of the mathematical Hopf map (Fig. 15c,d) [4] and relating experimental and theoretical closed-loop preimages of distinct $\mathbb{S}^2$-points [39]. The detailed structure of axisymmetric $\mathbf{m}(\mathbf{r})$ within the static Hopf soliton is depicted in Fig. 15e with the help of cross-sections orthogonal to the far-field $\mathbf{m}_0$ or containing it. Hopfions with different Hopf indices can be realized and co-exist in monodomain samples because they all can correspond to local or global free energy minima.

Minimization of free energy given by Eq. 8 predicts existence of 3D topological solitons in solid non-centrosymmetric ferromagnets [150] for experimental values of exchange energy and Dzyaloshinskii-Moriya constants $A_m$ and $D_m$. Like in the case of the chiral term in free energy for ferromagnetic colloidal systems [39], Dzyaloshinskii-Moriya term in Eq. 8, resembling the chiral term in the Frank-Oseen free energy functional Eq. 2, helps overcoming the stability constraints defined by the Derick theorem [151]. Inspired by experimental observations in CLCs and chiral ferromagnetic LC colloids, solid-state magnetic hopfions have been predicted to exist in nanodiscs, thin films and nanochannels of non-centrosymmetric magnetic solids with perpendicular surface anisotropy [150] (Fig. 17a,b), featuring closed-loop preimages of all $\mathbb{S}^2$ points, with each pair linked $Q$ times. Due to the field topology, the emergent magnetic field, defined by $(B_{em})_i \equiv \hbar \epsilon_{ijk} \mathbf{m} \cdot (\partial_j \mathbf{m} \times \partial_k \mathbf{m})/2$ with $\hbar$ being the reduced Planck's constant, of a solid-state elementary hopfion spirals around its symmetry axis with a unit flux quantum (Fig. 17c,d) [150]. Streamlines of $\mathbf{B}_{em}$, describing the interaction between conduction electrons and the spin texture, also resemble Hopf fibration [152]. This behavior of $\mathbf{B}_{em}$ mimics the topology of preimages for hopfions (Fig. 17e,f). It will be interesting to explore in future whether $\mathbf{B}_{em}$ in solid-state systems can also mimic behavior of preimages in high-Hopf-index hopfions, like the ones with $Q$=2 Solomon link topology (Fig. 17f). The capability of encoding 1, 0, 2, -1 and other states in the topological charges of 3D Hopf solitons in a chiral magnet can lead to data storage and other spintronics applications, with some of them already pursued in modelling [150,153]. While the stability of 3D solitons like hopfions has been always challenged by Derrick theorem [110,147,151–153], their experimental observation in chiral LCs and colloidal ferromagnets [39,110] offered insights that led to the predictions of such hopfions in magnetic solid-state materials [147,153], demonstrating the power of CLCs as model systems. The insight in this particular case is that the energetic stability of Hopf solitons is enhanced by the medium's chirality and that such topological objects can be hosted as stable or metastable structures in systems with Hamiltonians like the ones given by Eqs. 2 and 8 of chiral ferromagnetic colloidal LCs and solid-state magnets.

Hopf solitons in chiral nematics differ from the ones in vector fields of chiral magnets discussed above in that they are realized in the nonpolar field with the $\mathbb{S}^2/\mathbb{Z}_2$ order parameter space [110,154–156]. Rod-like molecules and **λ(r)** twist by $2\pi$ in all radial directions from the central axis to $\lambda_0$-periphery of the hopfions [110]. All $\mathbb{S}^2/\mathbb{Z}_2$-points for the nonpolar director have thus individual preimages in the form of two linked loops [110]. This is expected since the manifold $\mathbb{S}^2/\mathbb{Z}_2$ is effectively half of $\mathbb{S}^2$ and the smoothly vectorized version of the hopfion has all preimages of $\mathbb{S}^2$ in the form of individualized closed-loop regions. Although the **χ(r)** and **τ(r)** director fields in a hopfion are ill-defined for the far-field $\lambda_0$, one can still find locally twisting structure of **λ(r)** and apply the analysis of director fields accordingly. The result, unsurprisingly, is that the hopfion can be interpreted in terms of nonsingular disclination loops that we discussed above, which is also a loop of a 2D skyrmion. Not only elementary hopfions but entire zoos of $\pi_3(\mathbb{S}^2) = \mathbb{Z}$ and $\pi_3(\mathbb{S}^2/\mathbb{Z}_2) = \mathbb{Z}$ solitons exist [4,23]. The insights into the diverse interpretations of structural embodiments of topological Hopf solitons experimentally revealed by CLCs and chiral ferromagnetic colloidal LCs are useful for theoretical modelling and experimental discovery of such topological objects in other branches of physics.

### *8.2. Heliknotons*

Recently, another embodiment of the Hopf solitons, called "heliknotons", has been demonstrated in CLCs [32] that can be characterized by the already familiar triad of the orthonormal fields (Fig. 18a). Heliknotons are topological solitons with linked closed-loop **λ(r)**-field preimages (Fig. 18b) while their **χ(r)** and **τ(r)** contain half-integer singular vortex lines forming knots (Fig. 18c). Therefore, the heliknoton is a hybrid embodiment of both preimage and vortex knots [32]. These solitons embed into the helical background and form spontaneously after the transition from the isotropic to the CLC phase when an electric field is applied to a positive-dielectric-anisotropy chiral LC along the far-field helical axis $\chi(\mathbf{r})$. These structures comprise localized regions (depicted in Fig. 18b,c) of perturbed helical fields and twist rate, displaying 3D particle-like properties [32]. The inter-heliknoton interactions arise from sharing long-range perturbations of the fields and minimizing the overall free energy for different relative positions [32]. These interactions enable a plethora of crystals, including 2D and 3D low-symmetry and open crystalline lattices (Fig. 18) [32], with tunable crystallographic symmetries and lattice parameters [32]. 3D crystals of heliknotons emerge in samples of thickness $>4p$, when anisotropic interactions yield triclinic pedial lattices (Fig. 18d), whereas 2D crystals form in thinner samples (Fig. 18e-g). Besides the $Q=1$ elementary heliknotons, $Q=2$ and $Q=3$ topological solitons were observed as well [32], with preimages in the material field $\lambda(\mathbf{r})$ linked twice and three times, respectively. For $Q=2$ ($Q=3$) heliknotons, singular vortex lines in $\chi(\mathbf{r})$ and $\tau(\mathbf{r})$ form closed $5_1$ ($7_1$) knots co-located with the same knot of a meron in $\lambda(\mathbf{r})$. These and other heliknotons with even larger $Q$ can be ground-state and metastable structures [32], behaving like particles. However, unlike the atomic, molecular and colloidal crystals, heliknoton crystals exhibit giant electrostriction and dramatic symmetry transformations under <1 V voltage changes. They can potentially emerge in solid-state non-centrosymmetric magnets and ferromagnetic colloidal LCs with helical fields [4] and Hamiltonians similar to those of chiral LCs, as recently predicted theoretically [157].

By calculating the spatial variations of the triad of the three orthonormal directors {$\lambda(\mathbf{r})$, $\chi(\mathbf{r})$, $\tau(\mathbf{r})$} based on $\lambda(\mathbf{r})$, one can understand heliknotons as various knots and loops of $\lambda$ disclinations, which can exist as energetic minima depending on the LC parameters, boundary conditions, and the strengths of external electric field (Fig. 19) [44]. For example, a heliknoton in one of its embodiments exhibits three interlinked closed loops of the nonsingular λ lines for CLC with Frank-Oseen elastic constants $K_{33}=2K_{22}$, when each pair of the closed loops are linked once (Fig. 19a-c). On the other hand, when the bend elasticity of the CLC is reduced to $K_{33}=K_{22}$ (experimentally realizable [42–44,158]), a trefoil knot of a single loop (Fig. 19d-f), instead, emerges in the 3D topological soliton. Furthermore, the theory of the CLC disclinations within quaternion representation provides the insight that unlinking and relinking between the λ line loops are topologically allowed transformations [23,26,55,101], which is confirmed by observing a continuous configurational transformation between the three-ring-type and the single-knot-type heliknoton through careful adjustment of LC parameters in the computer simulations. The recent advent of novel LC systems with bend- and splay-relaxed elasticity [159–162] will provide a fertile ground for the exploration of stability of a variety of different 3D-localized solitonic structures under such different elastic anisotropy conditions. The numerical and experimental observations of continuous morphing of these localized field configurations in CLCs would serve as validation of these theories, whereas the physical behavior of such 3D topological solitons can be also enriched by the dynamics and out-of-equilibrium emergent transformations, which we will discuss in the next section for 2D skyrmions and torons.

## 9. Emergent dynamics of topological defects and solitons

CLCs can exhibit even richer interplay of topology and ordering under the out-of-equilibrium conditions. Topological defects and solitons in LCs often can be "activated" by supplying energy [64,65,146,163–167], just like this was done in the past with granular particles by shaking them [168–170], as an example. When supplied to CLC samples on macroscopic scales under well controlled conditions, the external energy can be converted into motion on the individual soliton or singular defect basis, which, in turn, can lead to various types of emergent collective behavior. Periodic pulses of applied field lead to squirming motions of individual skyrmions and torons [64,65] and then to collective schooling and orderly motions of hundreds-to-millions of such topologically-protected particle-like structures, with all motion directions selected spontaneously [163–166]. While in active nematics topological defects behave as active particles themselves [168,170,171], the topological solitons that we focus on here can be understood as active particle-like objects within an effectively passive medium [64,65,163–166]. Topological solitons move by invoking nonreciprocal rotational director dynamics, which could be paralleled with stadium waves (which move around the stadium without people leaving the seats). The LC soliton motions also resemble dynamics of topologically similar skyrmions in spin textures in solid-state chiral magnets that can move through rotations of spins within solid films with up to kilometer-per-second speeds [107,121–126]. The applied electric field, through its dielectric coupling with $\lambda(\mathbf{r})$, morphs a soliton (Fig. 20a-f) and elastic free-energy costs associated with this deformation tend to drive relaxation of $\lambda(\mathbf{r})$ back to the initial state that minimizes energy at zero applied field. The non-reciprocal nature of $\lambda(\mathbf{r})$-rotation in response to switching voltage on and off causes translation of solitons in lateral directions (Fig. 20g-i). Within each voltage modulation period $T_m$, the solitons are asymmetrically squeezed during the "field-on" cycle and relax to minimize the elastic free energy during the "field-off" cycle of $T_m$ (Fig. 20h,i) [64]. This periodic non-reciprocal asymmetric morphing of the localized $\lambda(\mathbf{r})$ resembles squirming in biological systems, albeit CLC solitons have no cell boundaries, density gradients or interfaces, so that the similarity is limited, mainly just in terms of the nonreciprocal character [65]. Like for active colloidal or granular particles [168,169,172–175], the energy conversion happens at the scale of individual particle-like solitons. Although the oscillating energy-supplying field is applied to the entire sample, its direction is not related to the emergent motion direction [64].

Elastic interactions between moving skyrmions emerge to reduce the free energy costs of $\lambda(\mathbf{r})$-distortions that each topological soliton induces, albeit without the dynamic $\lambda(\mathbf{r})$ reaching equilibrium because of the periodic voltage modulation and soliton motions [165]. These interactions are key to define the emergent collective behavior of skyrmions while they move, including the formation of schools of skyrmions with different types of internal clustering, as summarized in the diagram in Fig. 21. The nature of instantaneous interactions between continuously morphing solitons effectively changes within each $T_m$, but the overall collective behavior then arises from the cumulative effects of $T_m$-averaged instantaneous interactions. In presence of thousands-to-millions of skyrmions, the applied electric field initially induces random tilting of the director around individual skyrmions, so that their south-north preimage unit vectors $\mathbf{p}_i = \mathbf{P}_i/|\mathbf{P}_i|$ initially point along random in-plane directions. Individual skyrmions exhibit translational motions with velocity vectors $\mathbf{v}_i$ roughly antiparallel to their $\mathbf{p}_i$. With time, coherent directional motions emerge, with schooling of skyrmions either individually-dispersed or within various cluster-like assemblies [165]. Velocity and polar order parameters $S_v=|\sum_i^N \mathbf{v}_i|/(N v_s)$ and $S_p=|\sum_i^N \mathbf{p}_i|/N$ characterize degrees of ordering of $\mathbf{v}_i$ and $\mathbf{p}_i$ within the moving schools, where $N$ is the number of skyrmionic particles and $v_s$ is the absolute value of velocity of a coherently-moving school. Both order parameters increase with time from zero to ~0.9, indicating the emergence of coherent unidirectional motion of the particle-like topological solitons.

Similar to skyrmions, torons too can exhibit collective dynamics experimentally realized when perpendicular surface boundary conditions on surfaces confining the CLC are strong (Fig. 22a). A particularly interesting dynamic regime involves dense polycrystalline arrays of torons (Fig. 22b,c). An oscillating electric field **E** applied to a CLC with such polycrystalline arrangements of torons prompts motions of crystallites and lattice defects (Fig. 22d), showing behavior very different from that of skyrmions discussed above [164]. The external electric field **E** is again applied orthogonally to cell substrates and motions emerge along a spontaneously selected direction in a plane orthogonal to **E** (Fig. 22a-d). The crystallites of torons have different orientations of crystallographic axes of the quasi-hexagonal lattice relative to the average motion direction before and during motion (Fig. 22b-d). The temporal evolution of deformations of the complex director field upon turning $U$ on and off is not invariant upon time reversal, prompting lateral translations of torons, which synchronize to yield coherent motions of the crystallites of torons within quasi-hexagonal periodically deformed lattices (Fig. 22) [164]. Although the average direction of motion of toron crystallites is well defined, individual torons within the lattices execute rather elaborate "dancing-like" dynamics (Fig. 22e), where local translations in directions other than motion direction average out over longer periods of time. As a result, the primitive cells of crystals of torons are translated along the average motion direction (Fig. 22e) with velocities approaching a micrometer per second range (Fig. 22f). While there is no net displacement of toron lattices and both skyrmions and singular point defects within them at zero applied field (though thermal fluctuations of toron positions are present), this displacement becomes linear in time soon after the periodically oscillating voltage is applied (Fig. 22f). Numerical modelling and experiments reveal that this motion is accompanied by voltage-dependent lateral shifts of hyperbolic point defects and tilts/deformations of preimages as compared to those at $U=0$ (Fig. 22g,h). This electrically-powered self-shearing of torons is apparent when visualizing the south-pole preimages and lateral shifts of the singular point defects at opposite confining surfaces (Fig. 22h), as well as can be inferred from tracking point defects in bright-field microscopy (insets of Fig. 22f). Furthermore, all the preimages also rotate around an axis normal to the sample plane [164] (Fig. 22i). As voltage is effectively turned on and off within each period of square-wave modulation, toron's preimages rotate counterclockwise and clockwise (Fig. 22i), so that the director evolution that is manifested through such textural changes is not invariant upon reversal of time (Fig. 22i). Collective motions of crystallites of these solitons prompt fascinating evolution of grain boundaries and 5-7 defects [164]. The lattice defect motions are not generated by external stresses (like this would be the case, for example, during mechanical deformation of crystalline solids) but rather emerge from collective motions of crystallites of topological solitons within the material's interior, powered by the conversion of energy on an individual soliton basis. The collective dynamics of torons increases the orientational order, causes polar ordering of these asymmetrically sheared solitons, as well as leads to rather high velocity order parameters, which are all rather unexpected emergent effects [164]. These findings show that, being "activated" through supplying energy that is converted into motion locally, singular point defects, torons and skyrmions can emerge as active particles within the new breed of solitonic active matter [64,65,164,165,171,176–178]. Therefore, CLCs exhibit a great potential for revealing fundamental behavior of topological field configurations in and out of equilibrium.

## 10. Outlook and conclusions

Being highly experimentally accessible, CLCs not only provide a fertile ground for fundamental exploration of topologically nontrivial field configurations, but also serve as model systems for other branches of science. For example, we have seen above how, under different experimental conditions, they provide valuable insights into the nature of line defects in biaxial nematics and smectics and two-dimensional skyrmions and merons in chiral magnets. Studies of such topological objects in CLCs often took place long before similar topological field configurations attracted interest in other physical systems [93]. CLCs were also the first media where the three-dimensional Hopf solitons were experimentally discovered [4], which now also attract considerable and growing interest in solid-state magnets, ferroelectrics and optics [157,179,180]. However, the model-system potential of CLCs is far from being fully utilized. For example, although orthorhombic and monoclinic nonchiral nematics have been recently demonstrated in molecular-colloidal systems [46], their chiral counterparts still need to be demonstrated. While in the conventional CLCs only the $\lambda$ director is material and directors $\chi$ and $\tau$ are immaterial, defined by configurations of $\lambda$, low-symmetry CLCs could have two or all three such directors material. Such more complex CLCs could allow for stability of new types of multidimensional solitons and defects. For example, chirality could help energetically stabilize topological objects corresponding to the high-dimensional order parameter spaces of these systems, such as $SO(3)=\mathbb{S}^3/\mathbb{Z}_2$ of the chiral versions of monoclinic biaxial colloidal ferromagnets [181] and $SO(3)/D_2=\mathbb{S}^3/Q_8$ of the orthorhombic biaxial nematics [182]. What types of solitonic and singular knotted field configurations can be energetically stabilized in these low-symmetry analogues of CLCs? For example, $\pi_3(\mathbb{S}^3/Q_8)=\mathbb{Z}$ and $\pi_3(\mathbb{S}^3/\mathbb{Z}_2)=\mathbb{Z}$ topological solitons in biaxial chiral nematics and ferromagnetic LCs would be rather interesting analogues of the Skyrme solitons in high energy physics, but can they emerge as global or local free energy minima in these soft matter systems? What would be the fate of various solitonic and singular knots during monoclinic-orthorhombic, orthorhombic-uniaxial nematic and various other phase transitions involving these mesophases? Recently also becoming experimentally available, the helimagnetic and helielectric LC analogues of CLCs pose new questions about topological and energetic properties of defects and solitons in such systems [140,183–185]. The recent demonstration of 3D active nematics [186] also promises a variety of new opportunities in realizing various out-of-equilibrium vortices and solitons. For example, both active [186–193] and out-of-equilibrium passive LCs with "activated" dynamic defects [64,65,163–166] could reveal various analogues of topological instantons and nontrivial topological connectivity where dynamics of defects and topological solitons could even lead to formation of topological field configurations in a different class. For example, although topological objects in soft matter can be realized only in one-to-three-dimensional physical-configuration spaces, time in certain cases can be treated as an additional spatial dimension (say $\mathbb{R}^{3+1}$ for a 3D configuration space with certain special temporal dynamics and the corresponding $\mathbb{S}^4$-compactification) [60,194], so that an interesting question arises if topological objects predicted by the homotopy theory and labelled as $\pi_4(\mathbb{S}^3)=\mathbb{Z}_2$ and $\pi_4(\mathbb{S}^3)=\mathbb{Z}_2$ could be potentially realized in out-of-equilibrium soft matter systems.

      Particle-like topological solitons can be building blocks of exotic condensed matter phases too. Heliknotons have been shown to form various crystalline lattices, open and closed, including triclinic crystals (Fig. 18d-g) [32]. An important question is what it would take for the heliknotons to self-organize into solitonic LCs with only orientational ordering of the solitons. The interactions

between heliknotons already could be controlled between few $k_BT$ and hundreds of $k_BT$, as well as effective shapes could be controlled between nearly isotropic and highly elongated [32]. This could bring about hierarchical liquid crystallinity, where nanometer-long organic molecules form a chiral LC host medium that hosts heliknotons and then these micrometer-long topological solitons form yet another, solitonic LC on larger scales. What are the LC mesophases that can emerge in such systems? What types of new physical behavior and properties can arise because of these stable phases formed by topologically nontrivial solitonic field configurations? Technological needs and fundamental curiosity call for research to reveal such LC materials with different symmetries, topological diversity and varying combinations of order and fluidity.

From a more applied perspective, controlled patterning of defects in LC elastomers [195–197] can be extended to CLC skyrmions, hopfions and heliknotons, where topologically-protected nature of these field configurations can be utilized to produce well-defined localized mechanical responses, much like with singular defects [195–197], but now with a considerably larger inventory of possibilities. Topographic features at confining interfaces can be used to define and pattern spatial positions of various solitons through harnessing interactions mediated by CLC's orientational elasticity, much like colloidal particles in nematic LCs could be attracted by topographic features like pyramids [198]. On the other hand, optical effective refractive index patterns associated with various arrays of solitons with and without lattice defects can be utilized to generate tunable diffraction patterns and optical vortices in laser beams [116,117,199,200], where 3D topological solitons like heliknotons [32] within crystalline arrays may again allow for much needed reconfigurability in defining these diffractive elements and optical vortex generators. It will be interesting to explore how various linear and nonlinear optical interactions within CLCs can be exploited to use topological solitons in guiding laser beams of light and optical solitons like nematicons [201,202], as well as how these interactions can potentially enable practical applications in beam steering and telecommunications. CLC defects and topological solitons may also provide a platform for co-assembly of colloidal particles within these media, as already explored in several recent studies [99,203,204].

Acknowledgements

Authors are grateful to M. Bowick, J.-S.B. Tai, H. Zhao, C. Meng and B. Senyuk for helpful discussions. I.I.S. also acknowledges inspiring and insightful discussions with M. Kleman on topics of this review on multiple occasions since the first meeting, as well as is grateful to O. Lavrentovich for introducing him to both M. Kleman and CLC defects some 20 years ago. I.I.S. would like to thank his students and postdocs at the University of Colorado Boulder who have been working on research topics related to this review, including P.J. Ackerman, H. Mundoor, C. Meng, A.J. Hess, H.R.O. Sohn, J.-S.B. Tai, J.-S. Wu, B. Senyuk, J. Brewer, Z. Chen, J.S. Evans, M.G. Campbell, J. Bourgeois, T. Boyle, C.D. Liu, Q. Liu, A. Martinez, H.C. Mireles, Y. Wang, A.J. Funk, M.J. Laviada, I. Klevets, C. Lapointe, J. Giller, S. Park, O. Puls, D. Glugla, M.C.M. Varney, Q. Zhang, R. Voinescu, C. Twombly, G.H. Sheetah, A.J. Seracuse, M.B. Pandey, R.P. Trivedi, N.P. Garido, R. Visvanathan, O. Trushkevych. J.-S.W. and I.I.S. acknowledge the financial support of research on CLC colloids (the U.S. Department of Energy, Office of Basic Energy Sciences, Division of Materials Sciences and Engineering, under Award ER46921, contract DE-SC0019293 with the University of Colorado at Boulder) and topological solitons (the U.S. National Science Foundation grant DMR-1810513), examples of which have been highlighted within this review. I.I.S. also acknowledges the hospitality of the Kavli Institute for Theoretical Physics at the University of California, Santa Barbara, where he was working on this review during an extended stay.

**Figure Captions:**

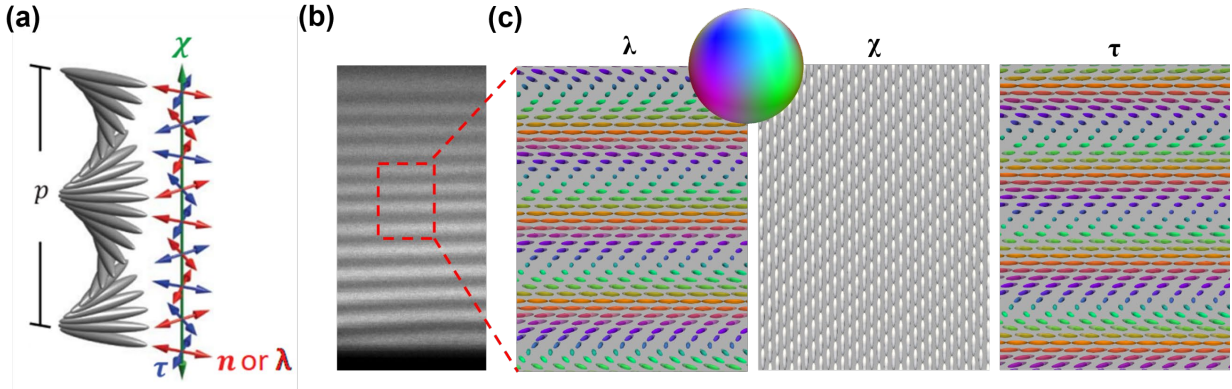

Figure 1. (a) A cholesteric LC with helical pitch $p$ and three orthonormal director fields $\lambda(\mathbf{r})$, $\chi(\mathbf{r})$, and $\tau(\mathbf{r})$ defined. (b) Topologically trivial helical state of a CLC observed experimentally using fluorescence confocal polarizing microscopy (FCPM). Bright stripes correspond to regions where $\lambda(\mathbf{r})$ directors are parallel to the polarization of an incident excitation laser beam, which in this case is perpendicular to the image. (c) Numerical visualizations of the three director fields corresponding to the region marked in (b). The directors are colored based on their orientations as shown in the order parameter space in the inset. The nonpolar nature asserts that the two opposite points on the sphere are labelled with the same color. The same color scheme is used for directors (ellipsoids) in all figures unless mentioned otherwise. Part (a) and (b) are reproduced with permission from Ref. [32] and Ref. [84], respectively.

|     | $\pi_1$ | $\pi_2$ | $\pi_3$ | $\pi_4$ |
|-----|---------|---------|---------|---------|
| $\mathbb{S}^0$ | 0 | 0 | 0 | 0 |
| $\mathbb{S}^1$ | $\mathbb{Z}$ | 0 | 0 | 0 |
| $\mathbb{S}^2$ | 0 | $\mathbb{Z}$ | $\mathbb{Z}$ | $\mathbb{Z}_2$ |
| $\mathbb{S}^3$ | 0 | 0 | $\mathbb{Z}$ | $\mathbb{Z}_2$ |
| $\mathbb{S}^4$ | 0 | 0 | 0 | $\mathbb{Z}$ |

Figure 2. Homotopy theory classification of singular and solitonic field configurations. The green, yellow and blue colors highlight different examples of topologically nontrivial field configurations discussed within this review, whereas the $\pi_3(\mathbb{S}^3)=\mathbb{Z}$ topological solitons (red) arise in high energy and nuclear physics models of subatomic particles.

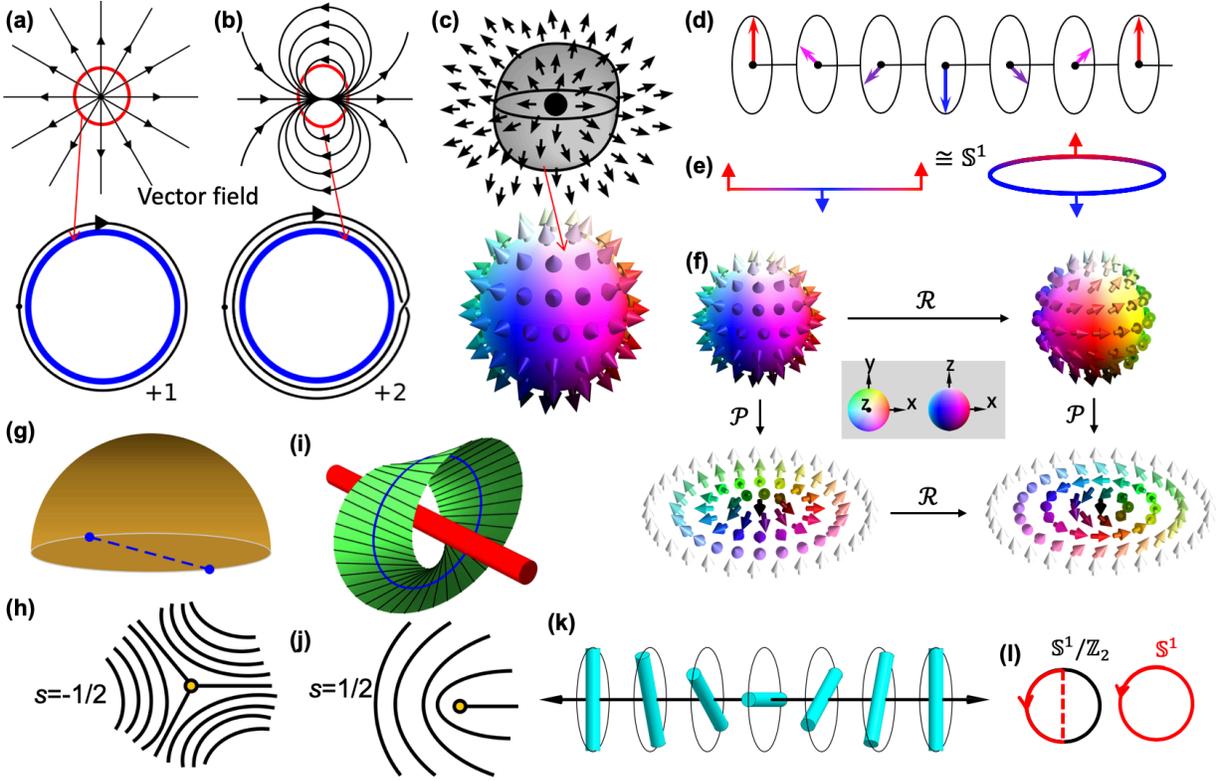

Figure 3. Topologically nontrivial structures of fields in condensed matter. (a,b) Examples of 2D $\pi_1(\mathbb{S}^1)=\mathbb{Z}$ singular defects classified by mapping the vector field from $\mathbb{S}^1$ surrounding the singularity to $\mathbb{S}^1$ order parameter space of vectors confined to 2D plane, with the order parameter space covered once in (a) and twice in (b), yielding the winding numbers. (c) An elementary +1 radial point defect representing a family of point singularities with integer-valued hedgehog charges labelled as $\pi_2(\mathbb{S}^2)=\mathbb{Z}$. (d) An example of $\pi_1(\mathbb{S}^1)=\mathbb{Z}$ topological soliton in the form of an elementary 1D solitonic wall with 360° unit vector rotation embedded in the uniform (vertical, pointing upwards) far-field background, which can be represented on $\mathbb{S}^1$, as shown in (e), where red-blue colors on $\mathbb{R}^1$ and $\mathbb{S}^1$ correlate with and depict vector orientations. (f) Skyrmions in $\mathbb{R}^2$ (bottom) can be mapped bijectively from field configurations in $\mathbb{S}^2$ (top) through stereographic projections ($\mathcal{P}$). The Neel-type (bottom-left) and Bloch-type (bottom-right) 2D skyrmions are related by a smooth rotation ($\mathcal{R}$) of vectors. The vector orientations are shown as arrows colored according to the corresponding points on the target $\mathbb{S}^2$ (inset). (Part f is reproduced with permission from Ref. [63]). (g) Schematic of the $\mathbb{S}^2/\mathbb{Z}_2$ order parameter space, with the diametrically opposite points of the circular base identified. (h-j) Half-integer defects in non-polar 2D $\boldsymbol{\lambda}(\mathbf{r})$, including wedge disclinations (h,j) that in 2D are characterized by opposite s=±1/2 winding numbers, and a twist disclination (i). In $\mathbb{R}^3$, there is only one type of topologically distinct disclinations different from a uniform state, with the topologically distinct states labelled $\pi_1(\mathbb{S}^2/\mathbb{Z}_2)=\mathbb{Z}_2$; local structures of defect lines like the ones shown in (h-j) can smoothly inter-transform one into another in 3D and correspond to a single, topologically equivalent state. (k) Twisted wall with 180° rotation of nonpolar $\boldsymbol{\lambda}(\mathbf{r})$ embedded in a uniform background can be compactified on $\mathbb{S}^1/\mathbb{Z}_2 \cong \mathbb{S}^1$. (l) Mapped director field of the twisted wall winds around the order-parameter space $\mathbb{S}^1/\mathbb{Z}_2$ once; since $\mathbb{S}^1/\mathbb{Z}_2 \cong \mathbb{S}^1$, 1D LC solitons are classified by $\pi_1(\mathbb{S}^1)=\mathbb{Z}$. Reproduced with permission from Ref. [4].

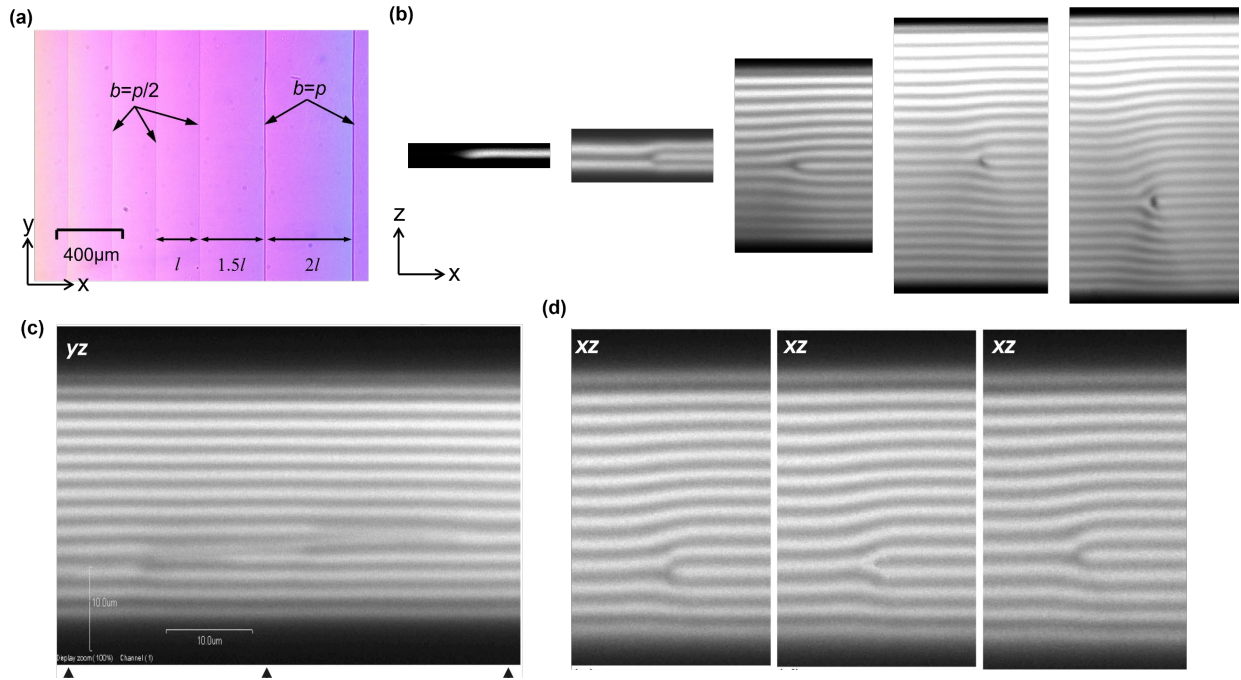

Figure 4. Edge dislocations and kinks in Grandjean-Cano wedges. (a) Polarizing microscopy textures showing an array of thin lines ($b=p/2$ dislocations) with interval $l=p/2\tan(\alpha)$ at the thinner part of the cell (left), and thick lines ($b=p$ dislocations) with spacing $2l$ at the thicker region (right). (b) A series of FCPM x-z slices in a Grandjean-Cano wedge showing a twisting disclination, $b=p/2$ dislocations, and a $b=p$ dislocation with increasing cell gaps. (c) A FCPM image along the glide plane (y-z plane) of a kink with height $p/2$ along a $b=p/2$ dislocation. The black triangles mark the positions of the vertical sections (x-z plane) in (d), showing the change in $\lambda(\mathbf{r})$ alignment along the kink. The helical pitch $p=5$ μm for all experiments, and polarization of the excitation beam is perpendicular to the images for (b-d). Reproduced with permission from Ref. [84].

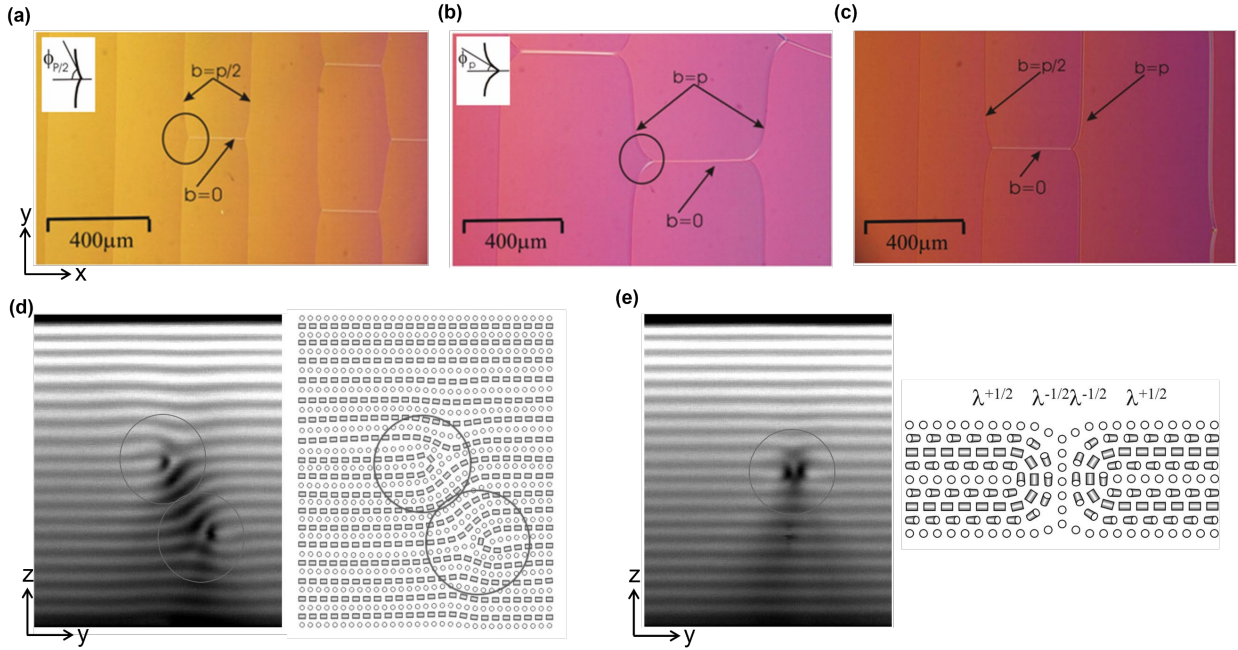

Figure 5. Dislocations with overall *b*=0 in Grandjean-Cano wedges. (a-c) Polarizing micrographs showing *b*=0 dislocations connecting *b*=*p*/2 dislocations (a), *b*=*p* types (b), and one of each type (c). The angle of defect lines at the joints are schematically shown in the insets in (a,b). (d,e) FCPM textures (left) and the corresponding director visualizations (right) taken perpendicular to *b*=0 dislocations revealing that the composing dislocations have the opposite Burgers vectors. The structure in (e) is known as Lehmann cluster. The polarization of the excitation beam is perpendicular to the images in (d,e). Reproduced with permission from Ref. [84].

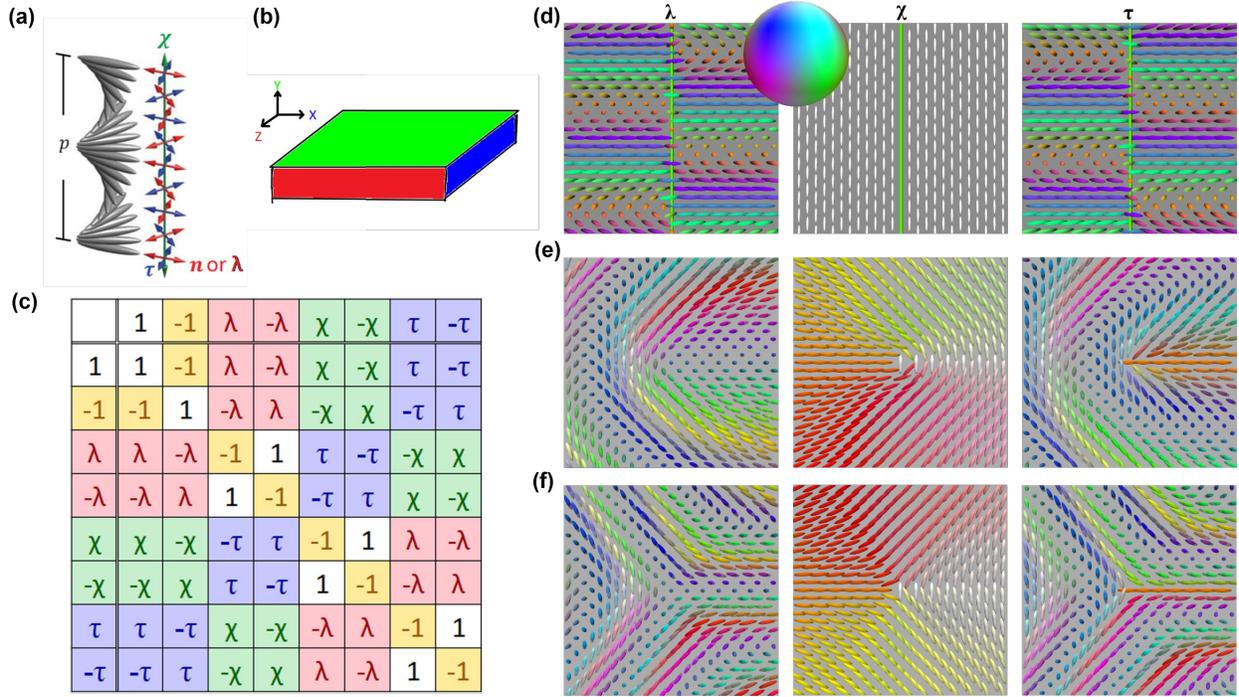

Figure 6. The topology of disclinations in CLCs. (a,b) The symmetry of the building blocks in cholesterics (a) and biaxial nematics (b) illustrated by the orthonormal frames {$\boldsymbol{\lambda},\boldsymbol{\chi},\boldsymbol{\tau}$} and {$\mathbf{x},\mathbf{y},\mathbf{z}$}, respectively. (c) The multiplication table of the quaternion group $Q_8$. All topologically stable disclinations in CLCs are characterized by one of the eight group elements, or one of the five conjugacy classes distinguished by colors. (d-f) Numerical visualizations of the $\boldsymbol{\lambda}(\mathbf{r})$, $\boldsymbol{\chi}(\mathbf{r})$, and $\boldsymbol{\tau}(\mathbf{r})$ director fields along a $\chi$ disclination (d), perpendicular to a $\lambda^{+1/2}$ line (e), or a $\lambda^{-1/2}$ line (f). The long green tube in (d) marks the core of the $\chi$ disclination. All directors are colored according to their orientations in the nonpolar order parameter space as shown in the inset in (d).

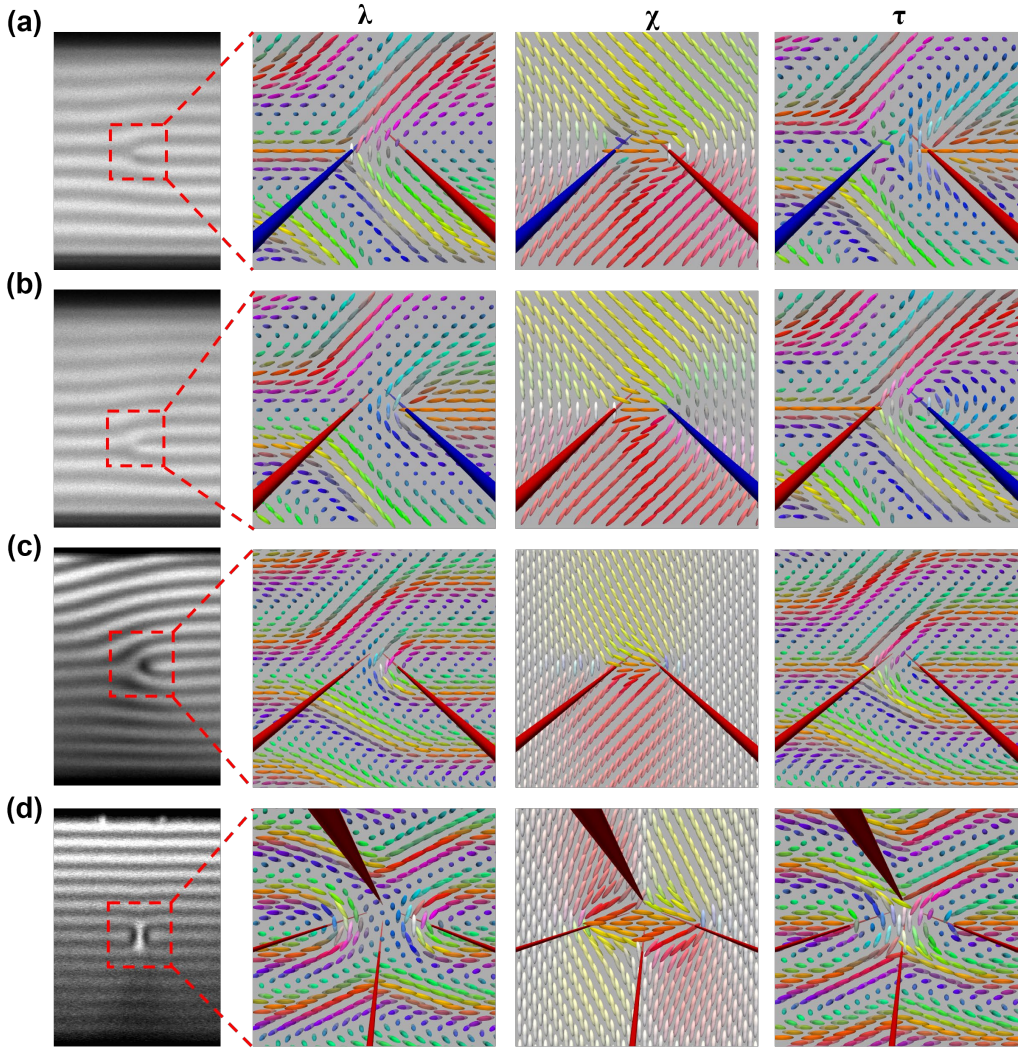

Figure 7. Dislocations as combination of disclinations. (a) Experimental FCPM image of a $\tau^{-1/2}$ $\lambda^{+1/2}$ dislocation (left) with the $\lambda(\mathbf{r})$, $\chi(\mathbf{r})$, and $\tau(\mathbf{r})$ director fields in the marked region reconstructed by numerical simulations (right). The positions of the $\tau^{-1/2}$ and $\lambda^{+1/2}$ disclinations are shown by the long blue and red tubes, respectively. (b-d) Similar visualizations for $\lambda^{-1/2} \tau^{+1/2}$ dislocation (b), $\lambda^{-1/2} \lambda^{+1/2}$ dislocation (c), and Lehmann cluster (d), with $\tau$ and $\lambda$ lines represented by blue and red tubes, respectively. FCPM images are reproduced with permission from Ref. [84] and simulations are carried out by numerical minimizing Eq. 11 with parameters matching those in experiments. Directors are colored according to their orientations.

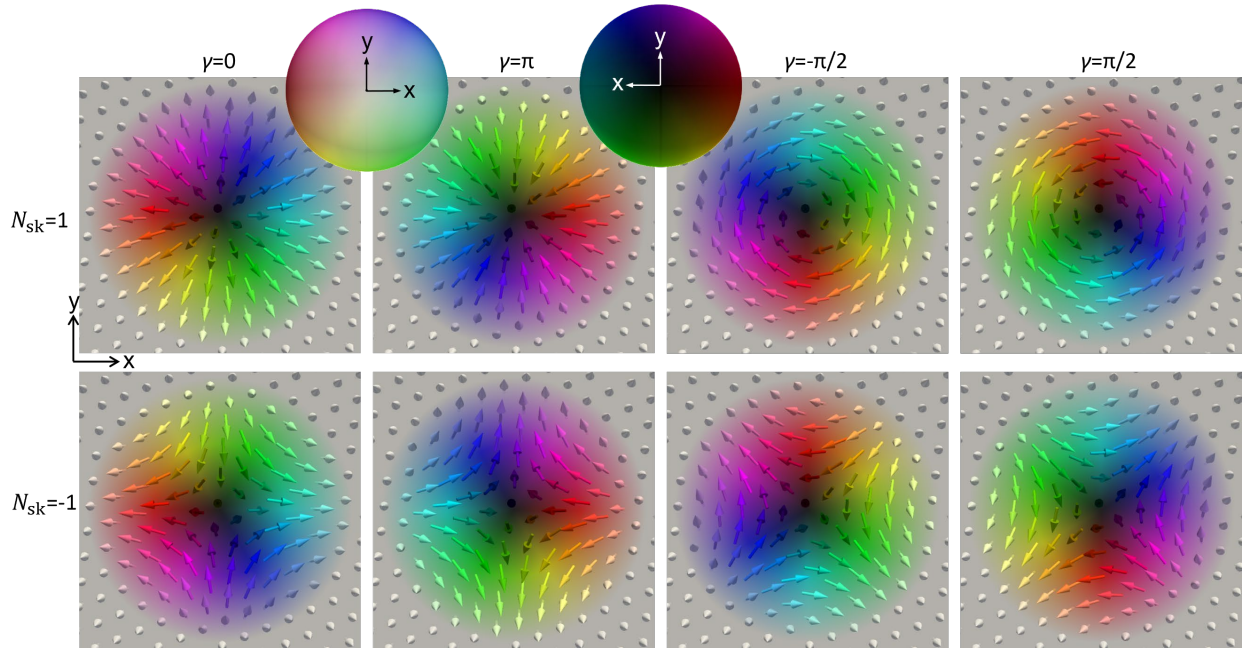

Figure 8. Full elementary skyrmions in magnetic systems. 2D Magnetic skyrmions characterized by the skyrmion number $N_{sk}$ and helicity $\gamma$. All arrows are colored according to their orientations in the order parameter space $\mathbb{S}^2$ shown in the insets.

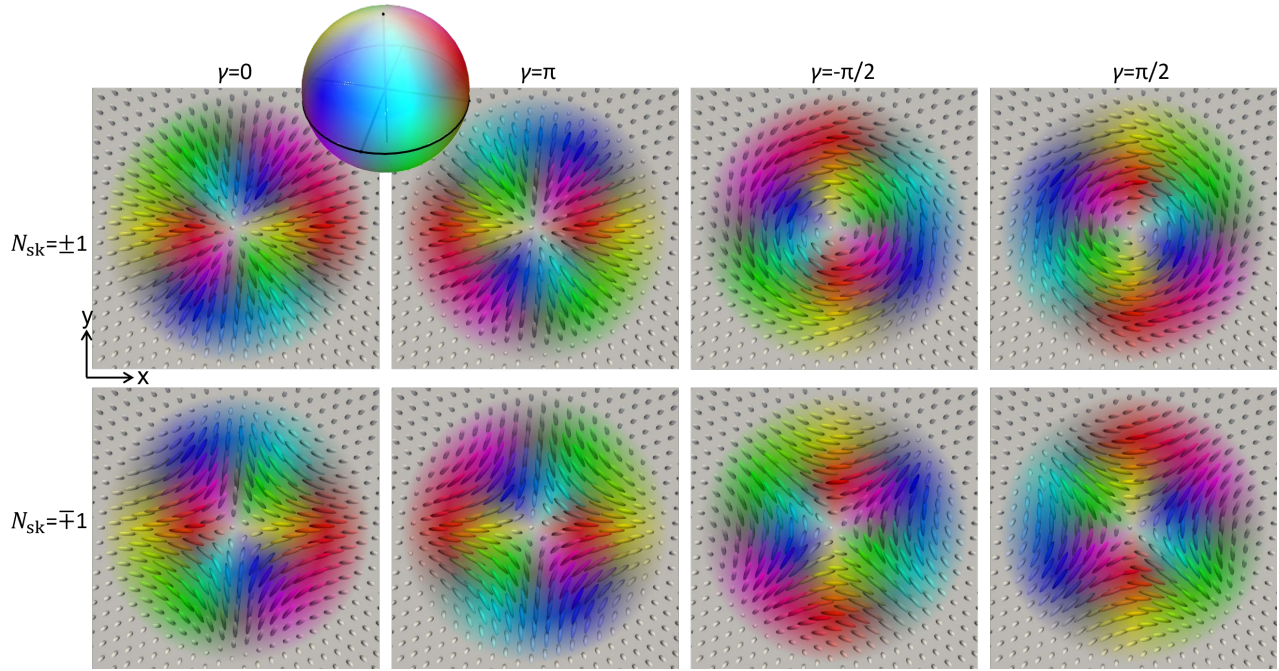

Figure 9. Full skyrmions in LC systems. Values of skyrmion number $N_{sk}$ and helicity $\gamma$ are marked for different 2D LC skyrmions. With the order parameter space of nonpolar LC directors being $\mathbb{S}^2/\mathbb{Z}_2$, we emphasize that each skyrmion wraps around $\mathbb{S}^2/\mathbb{Z}_2$ twice due to the nonpolar nature of the LC. Director orientations are colored according to their orientations.

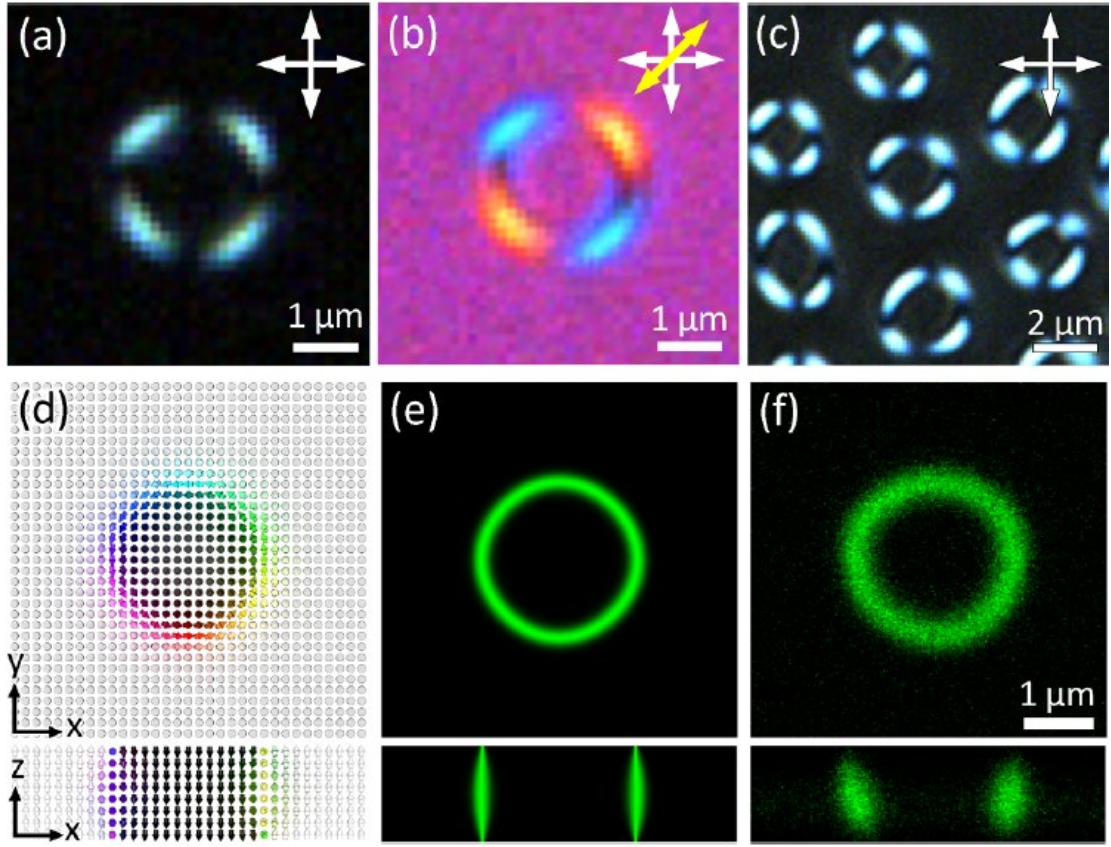

Figure 10. Experiments and modelling of 2D skyrmions. (a,b) Experimental polarizing optical micrographs of a 2D skyrmion obtained without (a) and with (b) a 530-nm phase retardation plate with its slow axis labelled by the yellow double arrow. (c) Experimental polarizing optical micrograph of an assembly of 2D skyrmions. (d) Midplane cross-sections of a numerically simulated stable skyrmion in a plane perpendicular (top) and parallel (bottom) to the far field. (e) Numerically simulated FCPM images obtained with circular polarization of a skyrmion in (d) in the midplanes perpendicular (top) and parallel (bottom) to the far field. (f) Experimental FCPM images obtained with circular polarization of a skyrmion in (a) and (b) in the midplanes perpendicular (top) and parallel (bottom) to the far field. The cell gap is $d=0.8$ μm and the CLC has pitch $p=1$ μm. Reproduced with permission from Ref. [150].

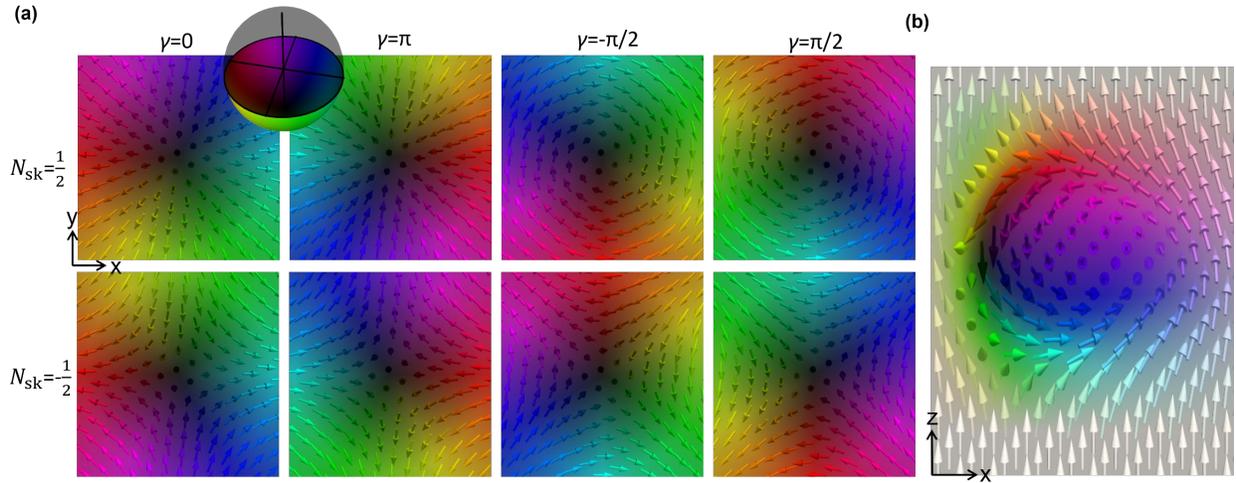

Figure 11. Magnetic fractional skyrmions (meron) with different values of skyrmion number $N_{sk}$ and helicity $\gamma$. (a) Each meron is mapped to, and colored according to, the half of the order parameter space, as shown in the inset. (b) The cross-section of a cholesteric finger of the second type in the vectorized director field is characterized by the overall $N_{sk}=1$, with the meron seen within it, in its center.

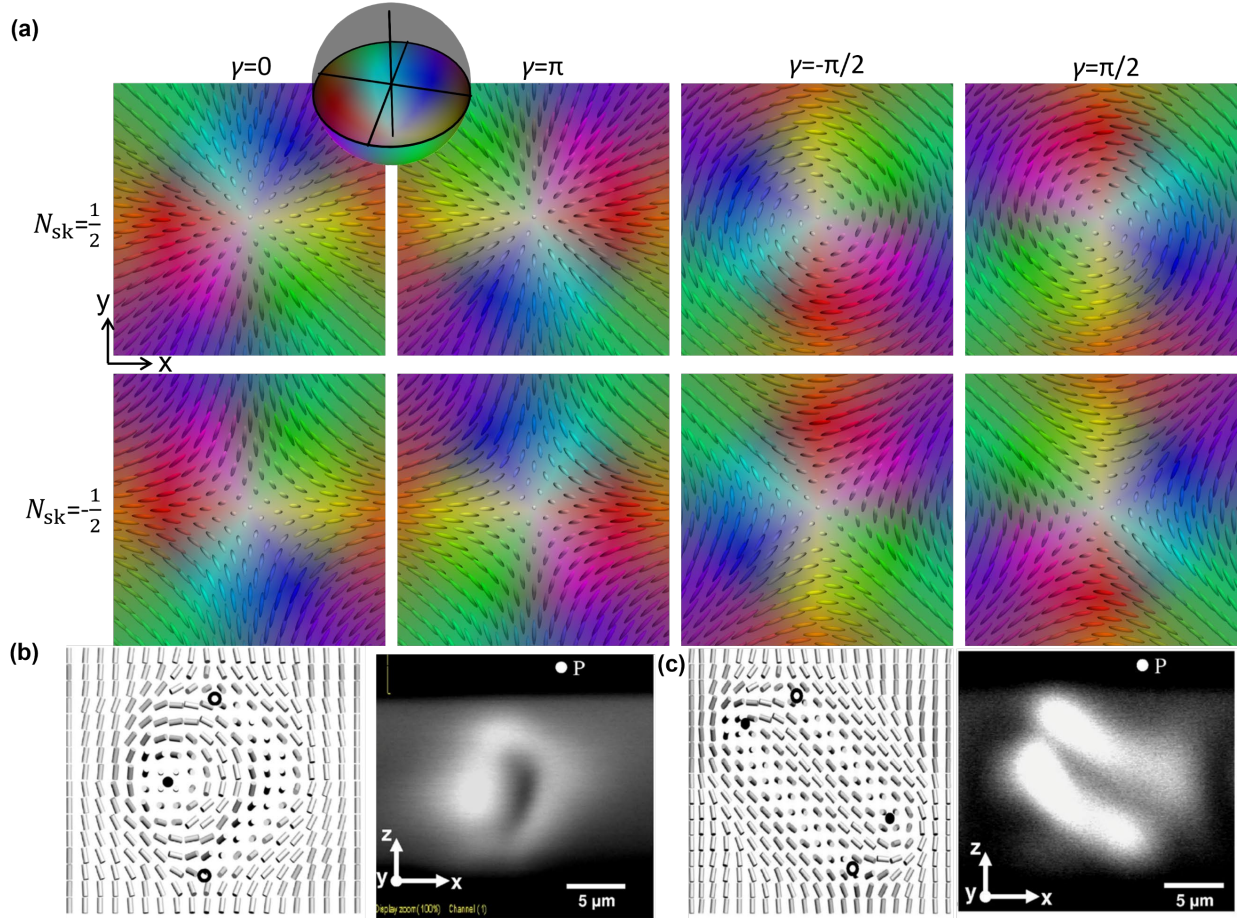

Figure 12. LC merons with different values of skyrmion number $N_{sk}$ and helicity $\gamma$ and as components of CLC fingers. (a) Each meron wraps around the order parameter space $\mathbb{S}^2/\mathbb{Z}_2$ once, as shown with the help of the color scheme depicting director orientations and relating it to the order parameter space in the inset. (b) Model (left) and cross-sectional FCPM image (right) of a cholesteric finger of the second type. In the model of director field, the half-skyrmion in the center and a quarter-skyrmions at the top and bottom are shown with filled and open circles, which add to unity and embed to a uniform far-field background. (c) Same as in (b), but for the cholesteric finger of the 1st kind, with filled and open circles showing quarter-skyrmions of opposite signs that add to zero and embed in a uniform background. Parts in (b,c) are reproduced with permission from Ref. [84].

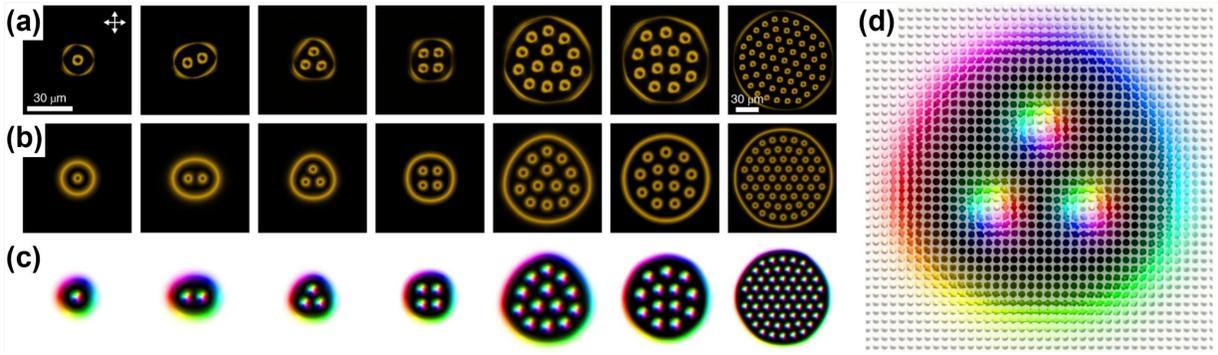

Figure 13. 2D skyrmions and skyrmion bags. (a) Polarizing optical micrographs of skyrmion bags with one-to-four antiskyrmions inside, two stable conformations of the bag with 13 antiskyrmions inside, and the bag with 59 antiskyrmions within it. (b,c) Computer-simulated counterparts of the skyrmion bags in (a). Crossed polarizers for (a,b) are marked by white double arrows in (a). (d) Close-up view of a computer-simulated bag with three antiskyrmions shown by colored, vectorized alignment field. Reproduced with permission from Ref. [107].

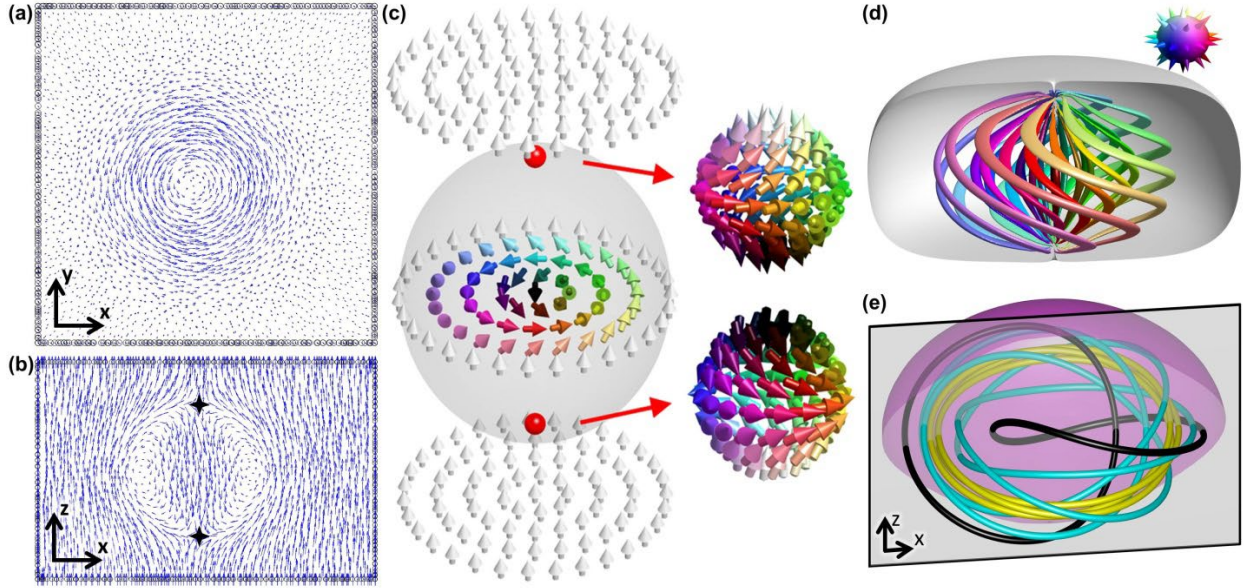

Figure 14. 3D structure and topology of elementary LC torons. (a,b) Computer-simulated cross-sections of an axisymmetric elementary toron shown in (a) plane orthogonal to $\lambda_0$ and (b) containing $\lambda_0$. (c) Elementary toron as a skyrmion terminating at the two point defects (red spheres) to meet uniform surface boundary conditions and match the topologically nontrivial skyrmion tube with the uniform far-field background of the 3D LC sample. Detailed field configurations on spheres around the point defects are shown as right-side insets. (Reproduced with permission from Ref. [63]). (d) Toron's preimages of $\mathbb{S}^2$-points (inset, shown as cones), with regions where preimages meet corresponding to point defects. (e) Closed-loop $\lambda(\mathbf{r})$-streamlines within the toron at different distances from its circular axis form different torus knots and links. $\lambda(\mathbf{r})$ directors are vectorized in (c,d) and before calculation of streamlines in (e). Reproduced with permission from Ref. [65].

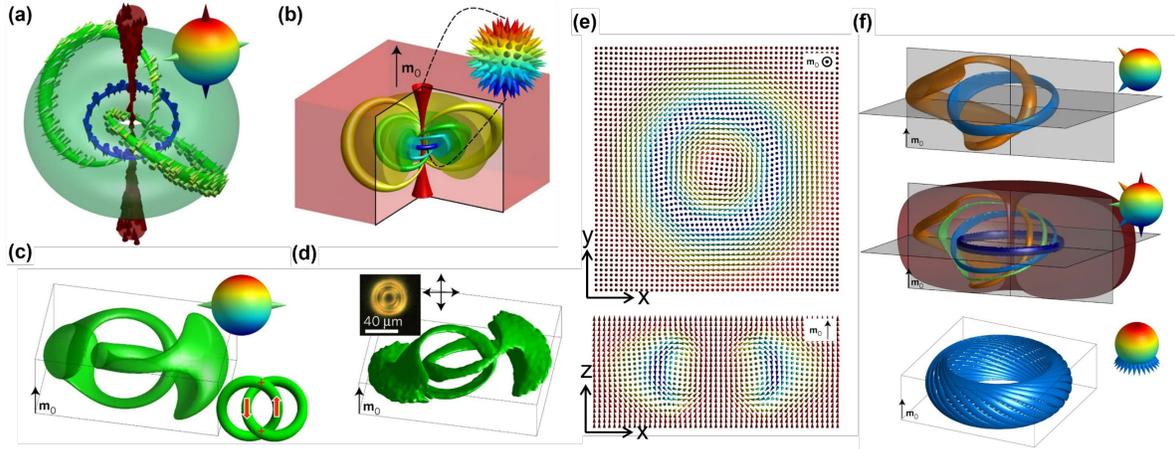

Figure 15. Hopfions in chiral colloidal ferromagnetic LCs. (a) Linking of hopfion's circle-like closed-loop preimages of points (cones) on $\mathbb{S}^2$. (b) Illustration of a Hopf map of closed-loop preimages of a hopfion embedded in a far-field $\mathbf{m}_0$ onto $\mathbb{S}^2$. (c, d) Computer-simulated and experimental preimages, respectively, of two diametrically opposite $\mathbb{S}^2$-points (cones) in the top-right inset of (c). Bottom-right inset in (c) shows signs of the crossings and circulation directions that determine linking of preimages. Inset in (d) is a polarizing optical micrograph of a hopfion. (e) Cross-sections of the hopfion taken in a plane orthogonal to $\mathbf{m}_0$ (top) and in a plane containing $\mathbf{m}_0$ (bottom), with the vector field shown using cones colored according to $\mathbb{S}^2$ shown in the insets of (a-c). (f) Linking of preimages with representative points on $\mathbb{S}^2$, including south-pole preimages corresponding to $\mathbf{m}_0$ confining all other preimages. Reproduced with permission from Ref. [39].

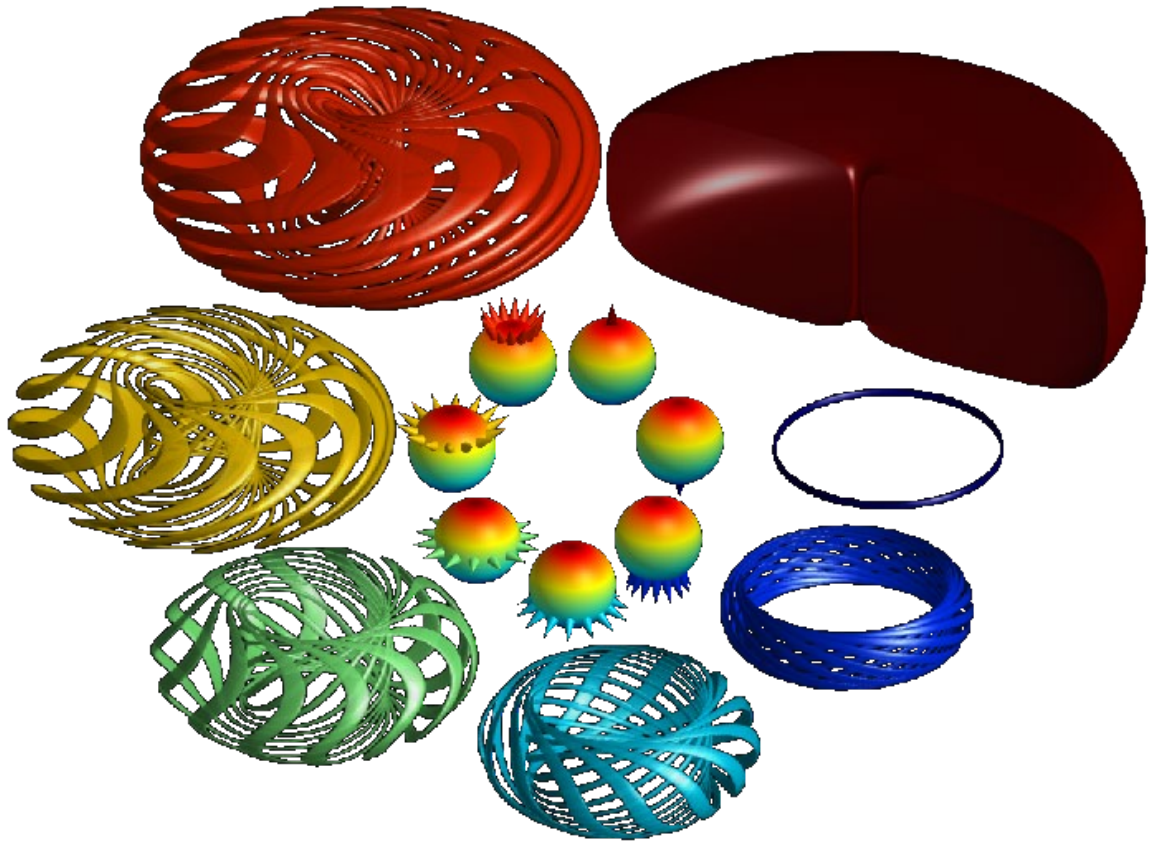

Figure 16. Tiling and linking of preimages of $\mathbb{S}^2$ points within a hopfion. Preimages of different azimuthal **m**-orientations tile into tori for the same polar angles, with the smaller tori nesting inside bigger tori; the largest torus contains the north-pole preimage in its exterior corresponding to $\mathbf{m}_0$ and the other preimages nested within its interior. Reproduced with permission from Ref. [4].

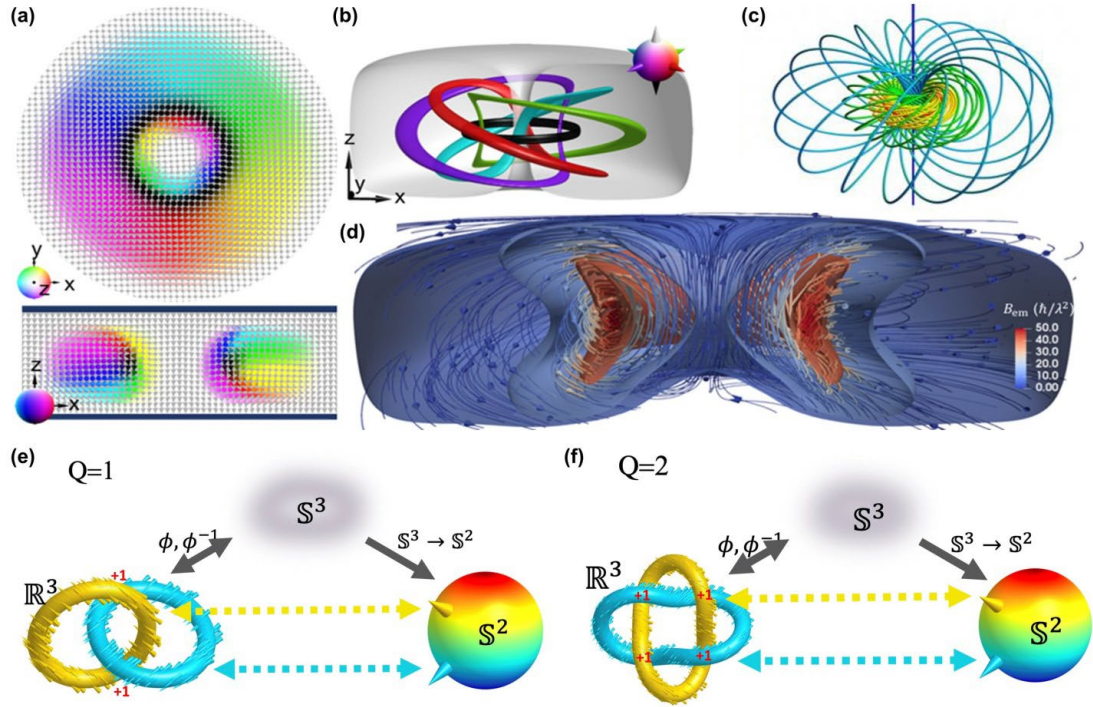

Figure 17. Hopfions in solid-state non-centrosymmetric magnets. (a) Cross-sections of the magnetization field within a hopfion in the plane perpendicular to $\mathbf{m}_0$ (upper) and that containing $\mathbf{m}_0$ (lower) in a magnetic solid material. Magnetization fields are shown with cones colored according to $\mathbb{S}^2$ (lower-left insets). In the x-z cross section, black stripes at the top and bottom indicate interfaces with boundary conditions achieved using perpendicular surface anisotropy and thin-film confinement. (b) Preimages of $\mathbb{S}^2$-points indicated as cones in the inset. Linking number of preimage pairs is consistent with the Hopf index $Q$=1. (c) Geometry and topology of Hopf fibration. (d) Visualization of the emergent magnetic field $\mathbf{B}_{em}$ by the isosurfaces of constant magnitude and streamlines with cones indicating directions. (a-d) Reproduced with permission from [42]. (e,f) For hopfions, preimages of $\mathbb{S}^2$ in $\mathbb{R}^3$ (and $\mathbb{S}^3$) form Hopf (e) and Solomon links (f) with linking numbers matching their $Q$=1 (e) and $Q$=2 (f) Hopf indices. Since direct ($\phi$) and inverse ($\phi^{-1}$) stereographic projections relate configurations on $\mathbb{S}^3$ and in $\mathbb{R}^3$ when embedded within $\mathbf{m}_0$, these solitons are characterized by $\mathbb{S}^3 \rightarrow \mathbb{S}^2$ maps, $\pi_3(\mathbb{S}^2)=\mathbb{Z}$ homotopy group and $Q \in \mathbb{Z}$; crossing numbers in (e,f) are marked in red. (e) and (f) reproduced with permission from Ref. [147].

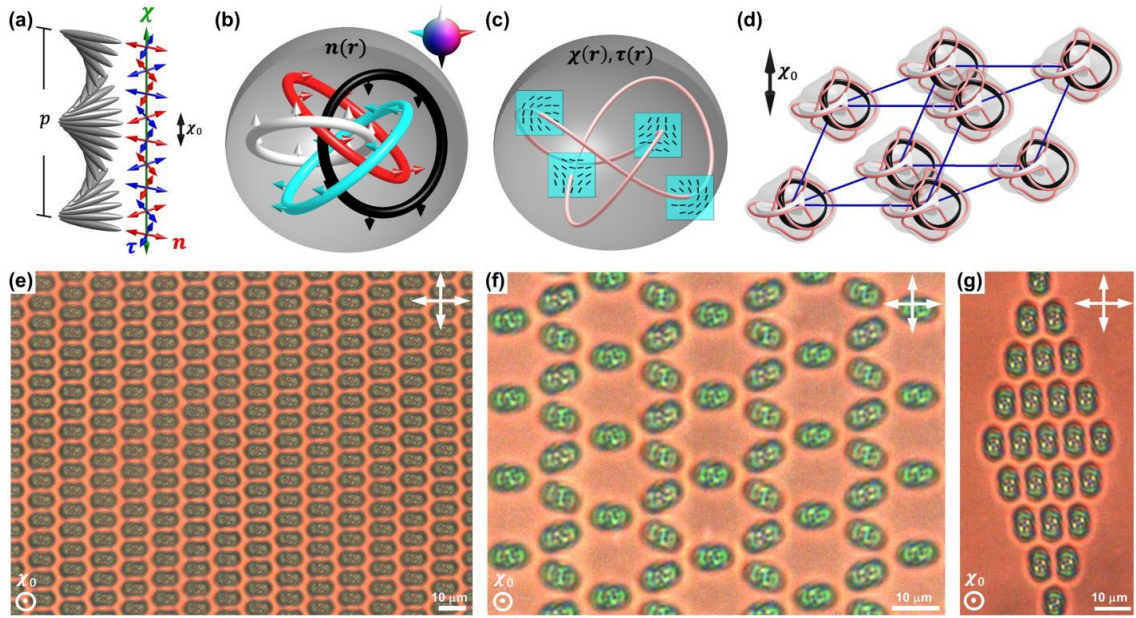

Figure 18. Topology and self-assembled crystals of heliknotons. (a) Helical field comprising a triad of orthonormal $\boldsymbol{\lambda}(\mathbf{r})$, $\boldsymbol{\chi}(\mathbf{r})$ and $\boldsymbol{\tau}(\mathbf{r})$. (b) Preimages of a heliknoton colored according to their orientations on $\mathbb{S}^2$ for vectorized $\boldsymbol{\lambda}(\mathbf{r})$ as shown in the inset. (c) Knotted co-located half-integer vortex lines in $\boldsymbol{\chi}(\mathbf{r})$ and $\boldsymbol{\tau}(\mathbf{r})$. Gray isosurfaces in (b,c) show the localized regions of the distorted helical background. (d) Primitive cell of a 3D triclinic heliknoton crystal. Isosurfaces (gray) of heliknotons with distorted helical background are colocated with both λ knots (red) and preimages of antiparallel vertical orientations in $\boldsymbol{\lambda}(\mathbf{r})$ (black and white). (e,g) closed rhombic and (f) open heliknoton lattices obtained at $U$=1.9 V and $U$=1.7 V, respectively. Reproduced with permission from Ref. [32].

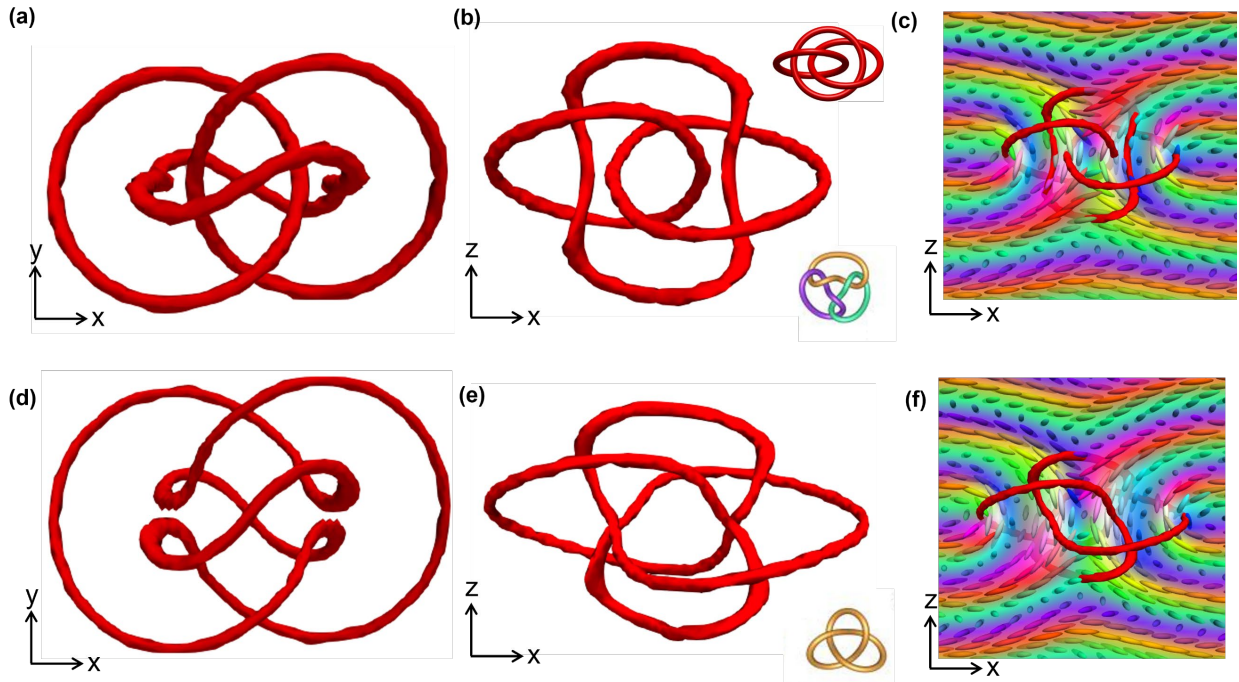

Figure 19. The loops and knots of λ lines in heliknotons. (a) Top view and (b) side view of three closed loops of λ disclinations within one embodiment of a heliknoton, with each pair of the loops linked. The simplified topological visualizations of the loops are shown in the insets of (b). (c) A vertical section of the $\lambda(\mathbf{r})$ director alignment for a heliknoton numerically stabilized under $U$=3 V, $d$=15 μm for CLC with p=5 μm. $K_{33}= 2K_{22}$. The long red tubes represent the geometry of the λ disclinations. (d-f) Similar visualizations but for a heliknoton simulated for $K_{33}= K_{22}$, showing a trefoil knot of the nonsingular λ line. Schematics in the lower right insets were generated using the KnotPlot freeware (https://knotplot.com), and the plotted directors adopt the same nonpolar color scheme as in Fig. 1.

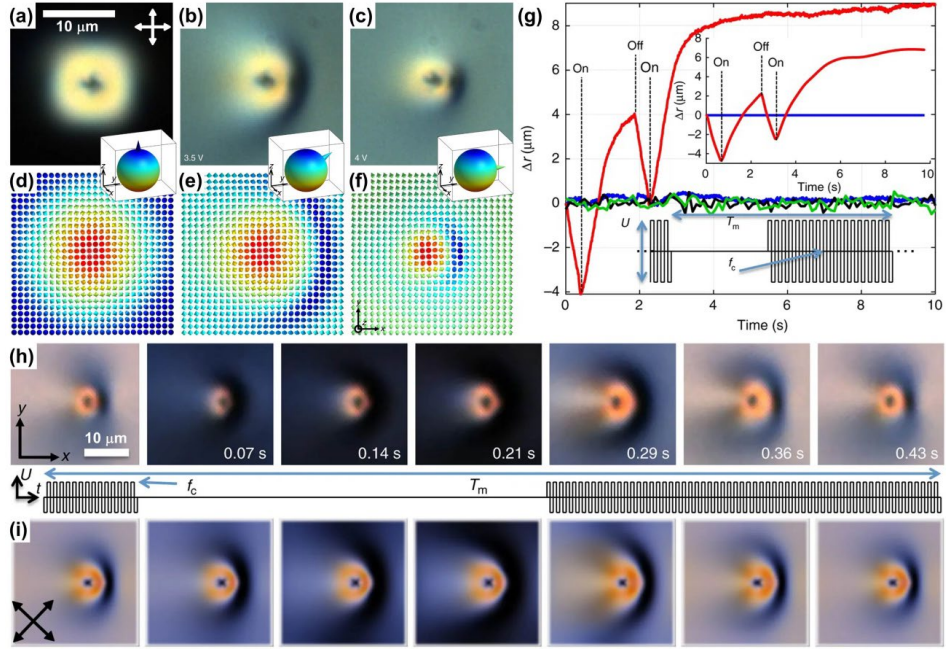

Figure 20. Translational skyrmion motions powered by an oscillating electric field. (a-f) Topology and electric switching (topology-preserving morphing) of 2D baby skyrmions: (a-c) polarizing optical micrographs of a 2D skyrmion at (a) no fields and (b, c) at voltages $U$ indicated on the images. Electric field applied to negative-dielectric-anisotropy LC is perpendicular to images. (d-f) Computer-simulated vectorized $\lambda(\mathbf{r})$ corresponding to (a-c), shown using arrows colored according to corresponding points on $\mathbb{S}^2$ (insets), with the far-field orientations depicted using cones. (g) Translation of a skyrmion in response to switching $U$ on and off, with corresponding computer simulated results shown in the top-right inset. The bottom inset illustrates the square waveform voltage driving with the carrier frequency $f_c = 1$ kHz and the modulation period $T_m$. Motion of the skyrmion is compared to that of a tracer nanoparticle at zero field (black solid line) and at $U=4$ V (green solid line). (h) Experimental and (i) computer-simulated polarizing optical micrographs of a skyrmion when moving along a vector connecting the south- and north-pole preimages (positive $x$). The schematic in the inset between experimental and computer-simulated micrographs shows the timing of turning $U$ on and off within the elapsed time equal to $T_m$, correlated with the micrographs in (h,i). Reproduced with permission from Ref. [64].

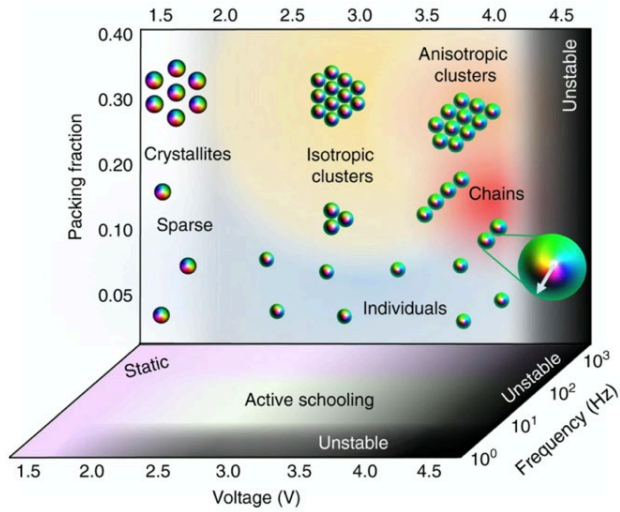

Figure 21. Schools of skyrmions characterized by a diagram of static and dynamic skyrmion assemblies and schools versus packing fraction, frequency $f_c$ and voltage $U$. The configurations shown in the insets are consistent across all $f_c$ at which skyrmions are stable. Reproduced with permission from Ref. [165].

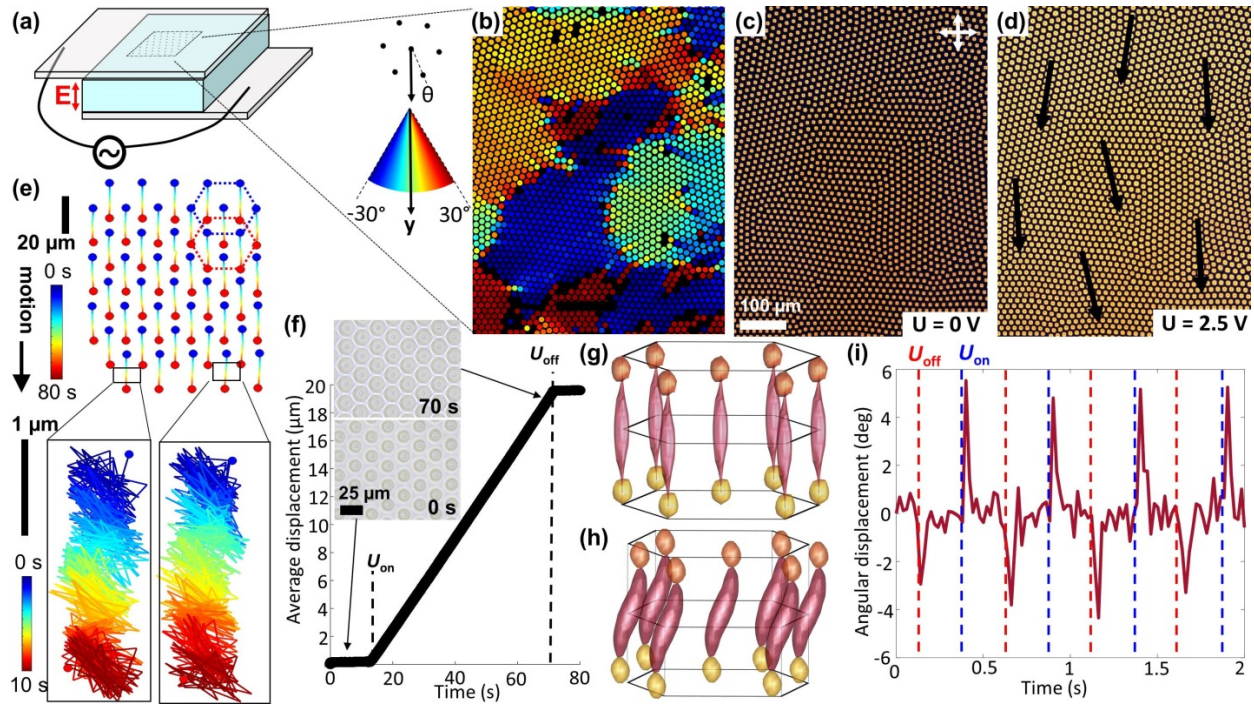

Figure 22. Dynamics of 2D crystallites of torons. (a) Schematic of crystallite motions powered by electric field orthogonal to the cell substrates. (b-d) Crystallites of torons colored according to orientations relative to the motion direction (b) and in polarizing micrographs at zero applied voltage (c) and at $U$=2.5 V (d); black arrows in (d) denote crystallite motions directions. Color scheme for visualizing crystallite orientations in (b) is shown in the inset between (a) and (b). (e) Trajectories of crystallite motions at $U$=2.5 V, $f_c$=10 Hz, progressively zooming in on the details of translations, colored according to elapsed time (with the maximum elapsed time marked in each part); dashed hexagons indicate the unit cell shift during motion, colored according to the color-coded timescale. (f) Average displacement of the hyperbolic point defects near confining substrate, analyzed with bright-field microscopy (video frames in insets). (g,h) South-pole preimages (magenta) and point defects (orange and yellow) of a hexagonal unit cell of torons shown (g) before and (h) during motion. (i) Nonreciprocal angular rotations of torons within crystallites upon voltage modulation, with the times of turning instantaneous voltage on and off marked by the blue and red dashed vertical lines. Reproduced with permission from Ref. [164].